\documentclass[a4paper,11pt]{article}
\usepackage{jcappub} % for details on the use of the package, please
\usepackage[T1]{fontenc} % if needed

\newcommand{\bx}{{\bf x}}

\def\be   {\begin{equation}}   \def\ee   {\end{equation}}

\title{\boldmath Entanglement in Cosmology}
\author[a]{K. Boutivas,}
\author[b]{D. Katsinis,}
\author[c,d]{G. Pastras}
\author[a]{and N. Tetradis}

\affiliation[a]{Department of Physics, University of Athens,\\Zographou 157 84, Greece}
\affiliation[b]{Instituto de F\'isica, Universidade de S\~ao Paulo,\\ Rua do Mat\~ao Travessa 1371, 05508-090 S\~ao Paulo, SP, Brazil}
\affiliation[c]{Institute of Nuclear and Particle Physics, NCSR `Demokritos',\\ Aghia Paraskevi 15310, Greece}
\affiliation[d]{Laboratory for Manufacturing Systems and Automation, Department of Mechanical Engineering and Aeronautics, University of Patras,\\Patra 26110, Greece}

\emailAdd{kboutivas@phys.uoa.gr}
\emailAdd{dkatsinis@phys.uoa.gr}
\emailAdd{pastras@lms.mech.upatras.gr}
\emailAdd{ntetrad@phys.uoa.gr}

\abstract{
We compute the evolution of the entanglement entropy for
a massless field within a spherical region throughout the inflationary period 
and the subsequent era of radiation domination, starting from 
the Bunch-Davies vacuum. 
In order to focus on the entanglement of modes that are directly
accessible to observations,  
we impose an ultraviolet cutoff set by the wavelength of the last mode that exited the
horizon at the end of inflation. The 
transition of each mode towards a squeezed state upon horizon exit during inflation and  
the additional squeezing when radiation domination sets in enhance the entanglement entropy.
Shortly after the transition to the radiation-dominated era,
a volume term develops and becomes the 
leading contribution to the entropy at late times, as is common
for systems lying in squeezed states. We estimate the magnitude of the entropy and 
discuss its interpretation in the light of the quantum to classical transition for
modes exiting the horizon during inflation. Our results raise the possibility that 
the quantum nature of weakly interacting fields, such as 
gravitational waves resulting from tensor modes during inflation,
may be detectable in today's universe. On the other hand, an 
observer with no knowledge of the degrees of freedom beyond the horizon would
interpret the entropy as thermal. From this point of view, the reheating after 
inflation would be a result of quantum entanglement.}

\begin{document}
\maketitle
\flushbottom

\section{Introduction}\label{introduction}

According to the inflationary paradigm, the large-scale structure of our universe
originated in vacuum fluctuations during inflation \cite{inflation1,inflation2,inflation3,inflation4,inflation5}.  
Quantum field fluctuations that were stretched beyond the horizon by the expansion were
transformed into classical stochastic fluctuations. After the end of inflation, they
re-entered the horizon at successive times depending on their wavelengths, and generated
the observed structure through the process of gravitational collapse of overdense regions. 

A rather obscure point in the above scenario is the quantum to classical transition
upon horizon exit. 
In inflationary cosmology a field fluctuation can be expressed in terms of 
momentum modes whose mode function involves a growing and a decaying term. 
After horizon exit, the mode function becomes
dominated by the first term and loses its oscillatory form (it freezes) \cite{physrep}. 
Moreover, the dominance of the
growing contribution causes the field and its conjugate momentum to commute.
As a result, the field after horizon crossing is viewed as a classical stochastic 
field, and its quantum expectation value is considered as the classical 
stochastic average.

Despite the simplicity of the above picture, 
it is important to keep in mind that the full quantum field and 
its conjugate momentum always obey the canonical commutation
relation. This is guaranteed by the presence of the decaying term 
in the mode function. In this sense, the field never loses its quantum nature. 
It has been argued that the observation of quantum properties of the field that may survive
until today is very difficult because of the enormous difference in the amplitudes 
of the growing and the decaying term, which would require a precision of 90 orders of
magnitude in the measurement of the momentum \cite{maldacenabell}. 
Devising experiments that could look for a violation of Bell's inequalities in 
the cosmological context seems like a formidable task \cite{maldacenabell,venninbell,vennin3}.

At the conceptual level, the question persists as to whether the quantum origin of
the fluctuations can be mirrored in certain quantities that probe beyond the classical
level. A very interesting example of such a quantity is the entanglement
entropy. In the simplest approach, one may consider the 
momentum-space entanglement 
between high and low-momentum modes,
such as between modes with physical momenta below and above
the Hubble scale $H$ \cite{momentumspace1,momentumspace2,momentumspace3,momentumspace4}.
However, this would vanish for a free field,
as long as the initial state can be written as a tensor product of states, one for
each momentum mode, as in the Minkowski vacuum.
Since each mode evolves independently, the reduced 
density matrix, resulting from some modes being traced over,
would be that of a pure state. A nontrivial result is obtained only in the 
presence of an effective coupling between different momentum modes, as in \cite{tripathy}.

The entropy associated with the entanglement 
between degrees of freedom
localized within two spatial regions separated by an entangling surface is
more promising. 
The calculation is more difficult, as one now has to trace over the degrees of freedom in the 
interior or the exterior. On the other hand, the reduced density matrix would not 
correspond to a pure state and the entanglement entropy would be nonvanishing even
in free field theory.
Explicit calculations of real-space entanglement entropy in flat space have been carried out 
for non-interacting or highly symmetric quantum field theories, and mostly in lower dimensions
\cite{Bombelli:1986rw,srednicki,callan,muller,casini0,casini1,casini2,pimentel,wilczek,korepin,cardy1,cardy2,Kanno:2014lma,Iizuka:2014rua,Kanno:2016qcc,stefan,Lohmayer,colas,vennin1}.
The entropy has interesting properties, such as
a dependence on the area of the entangling surface, which indicates a 
similarity with black hole entropy \cite{Bombelli:1986rw,srednicki}.

There are few results on the behaviour of real-space entanglement entropy in 
a cosmological setting \cite{pimentel,vennin2,Boutivas}. 
The explicit calculation of \cite{pimentel} focused on the
subleading logarithmic term, whose coefficient is universal. We are interested
instead in the leading term, for which we seek a physical interpretation.
Our approach follows closely the calculation of the entanglement entropy 
in flat (3+1)-dimensional spacetime in the seminal work by
Srednicki \cite{srednicki}. The original calculation 
focused on the leading term in the entropy, 
while the logarithmic correction was computed in \cite{Lohmayer}.
The basic formalism for the generalization of the calculation to the 
case of a  Friedmann-Robertson-Walker (FRW) background was developed in
\cite{Boutivas,Katsinis} and was applied to a toy (1+1)-dimensional model.
In this work we carry out the analysis for a (3+1)-dimensional cosmological background
that undergoes a transition from an inflationary era to a period of radiation domination. 

It must be mentioned also that the entanglement entropy can be derived through 
 the Ryu-Takayanagi proposal
\cite{ryu1,ryu2,review1} 
 in the context of the AdS/CFT correspondence \cite{adscft1,adscft2,adscft3}, and 
results have been obtained
in time-dependent backgrounds for theories that have gravitational duals 
\cite{tetradisgiataganas,tetradisgiantsos}.
Our aim here is to look directly at the details of the mechanism of entanglement in 
a cosmological setting for fields minimally coupled to gravity. For this, 
we perform an explicit calculation of the reduced density matrix 
in the case of a single massless scalar field, from which we deduce the
entropy. The effect of a non-minimal coupling has been explored in \cite{luongo1,luongo2}.

The main effect of the expansion on the 
quantum mechanics of the scalar field is that the field modes  
evolve from a simple oscillator ground state to a squeezed state 
\cite{Grishchuk}. There is a long history of studies on the connection between 
squeezing and entropy \cite{squeeze1,squeeze2,squeeze3,squeeze4,squeeze5,squeeze6,squeeze7,squeeze8}.
As we discussed above, 
the dominance of the growing term usually leads to the conclusion that 
the quantum properties of the field are not visible in late-time 
observations that focus on classical local quantities \cite{albrecht,classical1,classical2,classical3,classical4,classical5}.
However, the entanglement entropy may provide an exception, as 
it is a purely quantum, non-local quantity
that does not have a classical analogue. 
More specifically, the squeezing of canonical modes generically 
increases the entanglement between local
degrees of freedom and is expected to also increase the entropy. 

A conceptual question that arises when trying to interpret the entanglement entropy
is how to control the ultraviolet (UV) divergence that it displays. In flat 
(3+1)-dimensional spacetime
the entropy scales $\sim 1/\epsilon^2$, with $\epsilon$ a short-distance cutoff.
For the vacuum state this cutoff is intrinsically connected with the radius of the entangling
surface $R$, as they appear together in the leading term $\sim R^2/\epsilon^2$ that 
realizes the area law. The standard interpretation is that this term quantifies
the very strong entanglement between the short-distance modes on either side of the 
entangling surface. 

It is difficult to remove the divergent term through some kind of
renormalization procedure. On the other hand, if it is viewed as physical
there is an ambiguity in the identification of the length scale $\epsilon$. 
In a theory that includes gravity, one may
adopt the logic that $\epsilon$ must be of the order of the Planck length. 
However, there are fundamental difficulties with such an assumption. 
The huge difference between the Planck scale and macroscopic scales of interest would
mean that the entanglement entropy would always be dominated by the UV and would not
carry any interesting information about the long-distance physics. For an expanding
background the situation is more problematic. The continuous stretching of physical length
scales and the corresponding redshifting of frequencies imply that the UV is 
continuously replenished by new modes that emerge from sub-Planckian distances.
One can only make arbitrary assumptions about 
the state that such modes occupy, as well as their entanglement with the 
longer ones.

The quantum to classical transition upon 
horizon crossing during inflation can give a natural way to bypass the
above fundamental
issues. There is a certain mode of comoving wavenumber $k_s$ which
crossed the horizon at the end of inflation and immediately re-entered. Modes with 
wavenumbers $k>k_s$ remained subhorizon at all times and never went through the
process of freezing and the dominance of the growing term in the mode function. 
In this sense, they have always constituted vacuum fluctuations. Of course,
the value of $k_s$ is not fixed precisely, but the freezing of adiabatic modes 
occurs sufficiently
fast for $k_s$ to be determined up to a factor of order 1.
The modes with $k<k_s$ are the ones directly accessible to experiment and constitute the
observable universe. Even though they appear mostly classical, their quantum nature 
may still be visible in quantities such as the entanglement entropy. The advantage
of this logic is that the entanglement of interest is due to modes with 
wavelengths above a UV cutoff $\epsilon \sim 1/k_s$.

The absence of modes with $k>k_s$ can be justified only for very weakly interacting
fields, for which mode-mode coupling can be ignored. For interacting fields, all modes
are eventually excited. This does not result in an additional limitation 
on the type of fields that are possibly entangled today, as interacting
fields are expected to thermalize and lose quantum coherence anyway.
Our analysis is relevant for very weakly interacting fields as, for example, 
gravitational waves resulting from tensor modes during inflation. 
We treat them as freely evolving fields on the cosmological background, and 
we make the underlying assumption throughout the paper that their quantum coherence 
is not lost through some secondary
process during the whole evolution of the universe until today.

In order to obtain a feeling of the magnitude of $k_s$, we may use the fact that
a gravitational wave generated $N$ e-foldings before the end of inflation, with
frequency set by the approximately constant Hubble parameter $H$, is redshifted to 
a frequency $f$ today \cite{domcke}
\be
\ln\left(\frac{f}{10^{-18}{\rm~Hz}} \right)\simeq N_{\rm CMB}-N,
\label{frequency} \ee
where $N_{\rm CMB}\simeq 60$ indicates the number of e-foldings at which the modes of
the cosmic microwave background (CMB) exit the horizon. 
%The value of $N_{\rm CMB}$ depends on the energy scale of inflation. 
%The highest value of $N_{\rm CMB}$ consistent with observations 
%is $N_{\rm CMB}\simeq 60$. 
Modes that exit the horizon at the end of inflation 
($N\simeq 0$) would have a frequency today $f\sim 10^8 \rm{Hz}$,
which sets the cutoff in the spectrum of 
gravitational waves generated by inflation \cite{domcke}.
The corresponding wavelength is $\lambda_s \sim 1\rm{m}$. 
This is a scale much longer than any conceivable fundamental UV cutoff, but still 
much shorter than the Hubble radius today.

It is interesting also to note that a possible infrared cutoff $k_l$ may exist for the modes contributing to the entanglement entropy. Let us assume that
inflation started at some initial time, after some unspecified pre-inflationary era.
Even if modes that were superhorizon at that time
went through the process of freezing later on, 
their enhancement will be smaller than if they had grown
throughout inflation starting from below the horizon. The power spectrum for such
modes will be suppressed relatively to its value in the scale-invariant range.
It is then natural to expect
that these modes do not contribute significantly to the 
entanglement entropy. This, admittedly rather speculative, argument suggests that 
$k_l$ can be identified with the wavenumber of the mode that exited the 
horizon at the beginning of inflation.  

In the following we focus on the scenario in which the UV cutoff is identified with
the last mode that crossed the horizon during inflation. Its wavenumber satisfies 
$k_s/a_t\simeq H$, where $H$ is the constant value of the Hubble parameter during inflation, and
$a_t$ is the value of the scale factor at the 
transition from inflation to the era of radiation domination.
The UV regularization
is implemented by considering a discretized version of the free scalar theory on a lattice
of comoving spacing $\epsilon$. The highest mode has a comoving wavenumber $\sim 1/\epsilon$. 
Without loss of generality we can set $a_t=1$. This results in the
condition $k_s \simeq H$, or $\epsilon \sim 1/H$, where $H$ is the constant value of the
Hubble parameter during inflation. With these assumptions, 
the physical lattice spacing $\epsilon a$ never exceeds 
the physical Hubble radius during the whole evolution from the de Sitter (dS) phase 
 to the era of radiation domination (RD).

The paper is organized as follows:
In section \ref{formalism} we summarize the formalism that is needed in order to
compute the entanglement entropy in a time-dependent background. We describe the 
discretized version of the theory of a massless scalar field and the form of the 
wave function of the canonical modes. We also provide a concise summary of the 
method developed in refs. \cite{Boutivas,Katsinis} for the calculation of the
reduced density matrix and its eigenvalues, from which the entanglement entropy 
is deduced. In section \ref{numerics} we present the results of a numerical calculation
of the entanglement entropy in a time-dependent background that evolves from an 
inflationary era to radiation domination. In section \ref{conclusions} we give our
conclusions.

\section{Formalism}\label{formalism}

In this section we summarize the formalism that we employ for the calculation of
the entanglement entropy. We follow the approach of Srednicki \cite{srednicki}, which 
was generalized to the case of an expanding background in \cite{Boutivas,Katsinis}.
We refer the reader to these publications for the details. 

 \subsection{The discretized Hamiltonian} \label{discrHam}
We consider a free real scalar field in a FRW background, 
described by the metric 
\be 
ds^2=a^2(\tau)\left(d\tau^2 -dr^2-r^2d\Omega^2 \right)
\label{dsmetric} \ee
in terms of the conformal time $\tau$.
Through the definition $\phi(\tau,\bx)=f(\tau,\bx)/a(\tau)$,
the action of the field can be written as
\be
S=\frac{1}{2} \int d\tau\, d^3\bx\,\left(f'^2-(\nabla f)^2
+ \frac{a''}{a}f^2 \right),
\label{action1} \ee
where the prime denotes differentiation with respect to the conformal time $\tau$. 
The field $f(\tau,\bx)$ has a canonically normalized kinetic term. 
%i.e. its conjugate momentum is simply $f'(\tau,\bx)$.
As we consider
spherical entangling surfaces, it is convenient to define the spherical moments of the field and its momentum as
\begin{equation}
{f _{lm}} \left( r \right) = r \int {d\Omega \, {Y_{lm}}\left( {\theta ,\varphi } \right) f \left( {\bx} \right)} , \quad\quad
{\pi _{lm}}\left( r \right) = r \int {d\Omega \, {Y_{lm}}\left( {\theta ,\varphi } \right) \pi \left( 
{\bx} \right)} ,\label{eq:moments}
\end{equation}
where ${Y_{lm}}$ are real spherical harmonics. The radial coordinate can be discretized by
introducing a lattice of concentric spherical shells
with radii $r_j = j \epsilon$, where $j$ is an integer obeying $1 \le j \le N$. 
The radial distance between successive shells introduces an UV 
cutoff equal to $1/\epsilon$, while 
the total size of the lattice $L=N \epsilon$ sets an IR cutoff equal to $1/L$. 
By defining the discretized degrees of freedom as 
\begin{equation}
{f _{lm}}\left( {j\epsilon} \right) \to {f _{lm,j}} , \quad\quad
{\pi _{lm}}\left( {j\epsilon} \right) \to \frac{{{\pi _{lm,j}}}}{\epsilon}, 
\end{equation}
so they are canonically commuting,
we arrive at the Hamiltonian 
\be
H = \frac{1}{{2\epsilon}}\sum\limits_{l,m} {\sum\limits_{j = 1}^N {\left[ {\pi _{lm,j}^2 + {{\left( {j + \frac{1}{2}} \right)}^2}{{\left( {\frac{{{f_{lm,j + 1}}}}{{j + 1}} - \frac{{{f_{lm,j}}}}{j}} \right)}^2} + \left( \frac{{l\left( {l + 1} \right)}}{{{j^2}}}-\epsilon^2\frac{a''}{a}
 \right)f_{lm,j}^2} \right]} } .
\label{eq:Hamiltonian_discretized}
\ee
All dimensionful quantities in the problem can be expressed in units
of the {\it comoving} lattice spacing $\epsilon$. This is also the case for the 
time parameter $\tau$, as the combination $H \tau$ that determines the time evolution 
becomes a function of $\tau/\epsilon$, as is apparent from eq. (\ref{eq:Hamiltonian_discretized}). This normalization corresponds to setting $\epsilon=1$. 
There is a correspondence between $\epsilon$ and the comoving scale $k_s$ we discussed in the
introduction, i.e. $\epsilon \sim 1/k_s$. Similarly, we have that $L=N\epsilon \sim 1/k_l$. 
%We denote $\epsilon$ explicitly
%whenever the dependence on the UV cutoff is crucial. 

We shall analyze the scenario in which the evolution of the field starts during inflation
in a de Sitter (dS) phase and continues in the era of radiation domination (RD). 
We approximate the transition as instantaneous, neglecting the complications of reheating.
It is also possible to consider the transition to an era of matter domination (MD), 
either directly from the
dS phase \cite{Boutivas} or from the RD era. We have considered such scenarios as well, but they
lead to results that are very similar to the scenario we discuss in detail, albeit with 
much more complicated analytical expressions. 
Assuming that a transition from a dS to a RD background 
takes place at $\tau=0$, the scale factor takes the form 
\begin{equation}
a(\tau) = 
\begin{cases}
(1-H\tau)^{-1}, &\tau<0,\quad \text{dS},\\
1+H\tau, &\tau>0,\quad \text{RD},
\end{cases}
\label{scalefactors}
\end{equation}
so that
\begin{equation}
    a_\text{dS}(0)=a_\text{RD}(0)=1.
\end{equation}
The Hubble parameter is continuous through $\tau=0$. However, the 
effective mass term $-a''/a$ in eq. (\ref{eq:Hamiltonian_discretized}) is discontinuous.

The entanglement entropy between the interior and exterior of a sphere of
radius $R$ can be obtained from the density matrix of the harmonic system that corresponds
to the discretized field theory, via
tracing out the oscillators with $j \epsilon <R $ in order to compute the 
reduced density matrix for the degrees of freedom outside the entangling surface (or vice versa).
For the vacuum of the theory, the state of the
system of oscillators is the product of the `ground states' 
of the modes that diagonalize the Hamiltonian. 
The assumption of a Bunch-Davies vacuum implies that 
as `ground state' of a mode we must define the solution of 
the time-dependent Schr\"{o}dinger equation for this mode
that reduces to the usual simple harmonic oscillator ground state 
as $\tau \to - \infty$. 
The wave function of each mode depends on a linear combination of the 
various $f_{lm,j}$, i.e. the corresponding canonical coordinate. Since modes with 
different $l$ and $m$ indices do not mix, each eigenfunction actually involves one set of 
$(l,m)$. The effect of the expanding background is encoded in the term $\sim (a''/a)f_{lm,j}^2$, which is identical for all $(l,m)$. For a background that evolves from a dS to a RD era,
the discretized Hamiltonian for the free field takes the form
\be
H = \frac{1}{{2\epsilon}}\sum\limits_{l,m} {\sum\limits_{j = 1}^N {\left[ {\tilde{\pi} _{lm,j}^2 + \left( \omega_{lm,j}^2-{2\kappa}{\left(\frac{\tau}{\epsilon}-\frac{1}{H \epsilon}\right)^{-2}} \right)\tilde{f}_{lm,j}^2} \right]} } ,
\label{hamil}
\ee
 where $\tilde{f}_{lm,j}$ are the canonical coordinates 
 (linear combinations of $f_{lm,j}$), $\omega_{lm,j}$ the corresponding eigenfrequencies when the time dependence is neglected, and
 \begin{equation}
 \kappa = 
 \begin{cases}
 1, &\tau<0,\quad \text{dS},\\
 0, &\tau>0,\quad \text{RD}.
 \end{cases}
 \label{kappa10}
 \end{equation}

 At this point, we need to mention  a subtle issue concerning the cutoff imposed on the eigenfrequencies of the
  canonical modes of the above Hamiltonian. In the original calculation by Srednicki \cite{srednicki}, all discrete values of $l=0,...,\infty$ were taken into account. For 
  a given value of $\epsilon$, the sum over $m$ and $l$ appearing
  in eq. (\ref{hamil}) converges in the calculation of the entanglement entropy, so that 
  the UV divergence is well controlled by $\epsilon$. 
  In our approach, however, the fundamental issue is not simply one of regularization, but
  of the exclusion of all {\it physical} modes that remain subhorizon during the whole
  evolution. Allowing $l$ to take
  arbitrarily large values incorporates fluctuations in the tangential directions 
  that should have been excluded. In order to resolve this issue we have adopted the
  following strategy: We first determine the maximal eigenfrequency $\omega_{\rm max}$
  in the $l=0$
  sector, which is of order $1/\epsilon$. In all other $l$ sectors, we exclude the 
  eigenfrequencies that exceed $\omega_{\rm max}$, and the corresponding canonical modes.
  The resulting entanglement entropy still has a UV cutoff set by $\epsilon$, but a
  difference may appear in non-universal contributions. In particular, the leading term
  $\sim R^2/\epsilon^2$ for a static background will have a different coefficient than
  the one computed in \cite{srednicki}. Actually, the coefficient turns out to be smaller than
  in \cite{srednicki}, as an infinite number of modes are excluded. The procedure may
  be viewed as a different regularization of the entanglement entropy. However, our
  interpretation is that it provides a result that incorporates only the modes that
  went through the physical process of freezing upon horizon exit. 
  
  The cutoff procedure is also related to the presence of a new scale in the calculation
  of the entropy,
  set by the Hubble constant $H$ during inflation. In the introduction we argued that 
  the relevant scale of highest frequency for the problem is the one with comoving 
  wavenumber $k_s$, such that the mode exits the horizon at the end of inflation
  and immediately re-enters in the RD era. For this mode $k_s/a_t=H$,
  where $a_t$ is the value of the scale factor at the transition from the dS to the RD era. 
  We have set $a_t=1$ without loss of generality, 
  while the shortest
  mode that we keep in the calculation has an eigenfrequency set by $\epsilon$. This 
  leads to the identification $\epsilon \sim 1/H$, which we employ in the 
  next section.

 Of course, one could ignore the issue of mode freezing and treat
  $H\epsilon$ as a free parameter. 
  In this case, the scale $1/\epsilon$ can be viewed as a coarse-graining scale in a
  Wilsonian procedure of integrating out high-energy modes. As the momentum modes of a free
  field are decoupled, the integration would result again in a free theory, albeit with fewer
  modes. Even though this logic seems appealing in momentum space, it is less 
  transparent in real space. 
 The degrees of freedom in this case
  are the field values at every point in the discretized space (the lattice), 
  which are coupled through the
  kinetic term. Varying the lattice spacing essentially defines a new entropy that depends
  on an arbitrary UV scale. 
  Moreover, $H\epsilon \gg 1$ results in a discretized
  version of the field theory such that the characteristic
  length scale of the gravitational background can become smaller than the 
  physical lattice spacing during part of the evolution.
  On the other hand, 
  $H\epsilon \ll 1$ results in the dominance of the UV fluctuations and a result
  for the entanglement entropy identical to that in static space at all times.
  The natural choice $H\epsilon \sim 1$ that we advocate avoids the above complications.

\subsection{The wave function of the canonical modes}

The quantization procedure treats the field as a collection of 
quantum harmonic oscillators. 
	In order to make the analogy with the language of quantum mechanics clearer, we follow the notation 
	of previous work \cite{srednicki,Boutivas,Katsinis} and
	use the variable $x$ instead of
	$f$.  It must be kept in mind that it is
	the field value that is treated as a harmonic oscillator. In this section the variable $x$ has nothing to do with spatial coordinates.
%	In position space, $f$ is a function of $\mathbf{{x}}$.
The Hamiltonian (\ref{hamil}) becomes diagonal when expressed in terms of canonical modes,
which must be put at their `ground state' in the vacuum of the theory. The form 
of the Hamiltonian implies that we 
need the `ground state' eigenfunction of the harmonic oscillator with 
a time-dependent frequency of the form\be
\omega^2(\tau)=\omega^2_0-\frac{2\kappa}{(\tau-1/H)^2}.
\label{tdmass} \ee 
We set $\epsilon=1$ for simplicity in the expressions. Since the Hamiltonian has explicit time dependence, this `ground state' is not an eigenstate of the Hamiltonian, but rather a solution of the time-dependent Schr\"{o}dinger equation that reduces to the ground state of the simple harmonic oscillator with constant frequency $\omega_0$, as $\tau \to - \infty$.

A solution of the time-dependent Schr\"{o}dinger equation for an oscillator with
frequency given by eq. (\ref{tdmass}) can be obtained  
in several steps \cite{guerrero}, following the Lewis-Riesenfeld
method \cite{lewis1,lewis2}.
First one must find a solution $b(\tau)$ of the Ermakov equation 
%\cite{ermakov}
\be
b''(\tau)+\omega^2(\tau)b(\tau)=\frac{\omega^2_0}{b^3(\tau)},
\label{ermakoveq} \ee 
with boundary conditions appropriate to the problem at hand.
In terms of $b(\tau)$, the solution 
of the time-dependent Schr\"odinger equation 
is given by
\be
F(\tau,x)=\frac{1}{\sqrt{b(\tau)}}\,\exp\left({\frac{i}{2}\frac{b'(\tau)}{b(\tau)}x^2}\right)\,
F^0\left(\int \frac{d\tau}{b^2(\tau)},\frac{x}{b(\tau)}\right),
\label{solschrod} \ee
where $F^0(\tau,x)$ is a solution of the standard simple harmonic oscillator with
constant frequency $\omega_0$, namely a linear combination of the wave functions
\be
F^0_{n}(\tau,x)=\frac{1}{\sqrt{2^n n!}}\left(\frac{\omega_0}{\pi} \right)^{1/4} 
\exp\left(-\frac{1}{2}\omega_0 x^2 \right) 
H_n\left(\sqrt{\omega_0}x \right)
\exp \left(-i\left(n+\frac{1}{2} \right)\omega_0\tau \right),
\label{quantharm} \ee
with $H_n (x)$, $n=0,1,2,...$ the Hermite polynomials.
%Thus, the problem is reduced to solving eq. (\ref{ermakoveq}) with appropriate 
%boundary conditions. 

The solutions (\ref{solschrod})
are not energy eigenstates, as the problem has an explicit time dependence.
The `ground state' must be selected 
through appropriate boundary conditions. 
We require that the solutions are reduced to the standard wave functions of the harmonic oscillator
for $\tau\to -\infty$, consistently with the selection of the Bunch-Davies vacuum for the
field theory. For the `ground state', we have $F^0(\tau,x) = F^0_0(\tau,x)$, while the function 
$b(\tau)$ must satisfy $b(\tau)\to 1$ for $\tau\to-\infty$. At the time $\tau=0$
corresponding to the transition to the RD era, we impose the continuity of the wave function.
This fixes both $b(0)$ and $b'(0)$ and determines uniquely the form of $b(\tau)$ for $\tau>0$. We refer the reader to
\cite{Boutivas} for the details of this straightforward 
calculation.
We list here the expressions for the function $b(\tau)$ in the dS and RD eras, which
determine the `ground state' for our problem:
\begin{eqnarray}
b^2_{\rm dS}(\tau)=&1+\frac{1}{\omega_0^2\left( \tau-\frac{1}{H} \right)^2},
\quad\quad \quad \quad \quad \quad \quad \quad \quad \quad \quad\quad \quad \quad \quad \quad
\tau&<0,
\label{bdS2sf} \\
  b^2_{\rm RD}(\tau)=&
  1+\frac{H^4}{2\omega_0^4}
  +\left(\frac{H^2}{\omega_0^2}-\frac{H^4}{2\omega_0^4} \right)
  \cos(2\omega_0 \tau)
  + \frac{H^3}{\omega_0^3}\sin(2\omega_0 \tau),
  \quad 
  \tau&>0.
  \label{bRD2f} 
  \end{eqnarray}
It can be seen easily that $b_{\rm dS}$ tends to 1 for $\tau\to -\infty$, while
$b_{\rm dS}$ and $b_{\rm RD}$, as well as their derivatives, are continuous at $\tau=0$.
%We have checked explicitly that the functions (\ref{solschrod}) satisfy the time-dependent Schr\"odinger equation.

\subsection{The calculation of the entanglement entropy}

In this work we calculate the entanglement entropy following the original approach
by Srednicki \cite{srednicki}. A huge advantage of this approach is that it provides the reduced density matrix 
as an intermediate result. This matrix contains all the information about entanglement, and the entropy can be derived from its spectrum.
It must be kept in mind that the inverse process is not possible, 
since the entanglement entropy provides only partial information about entanglement.

Srednicki's method treats the field as a collection of coupled harmonic oscillators.
%$\tilde{f}(\mathbf{{x}})$. 
When the state of these oscillators is specified, the determination
of the density matrix is straightforward. The next step is to 
trace out some of the degrees of freedom, corresponding 
to the field values on one side of an entangling surface, in order to 
derive the reduced density matrix. 
The difficulty of this task depends on the state in which the overall system lies. The diagonalization of the Hamiltonian decouples this system by describing it in
terms of the field canonical modes.
In the original calculation \cite{srednicki} the field is considered in 
flat spacetime and is assumed to be at its ground state, i.e. each decoupled canonical mode is at its ground state. This greatly simplifies the calculation, 
as the ground state is Gaussian, and tracing out degrees of freedom can be 
carried out via the evaluation of Gaussian integrals. In our case
the field theory is considered in an expanding background. There is an effective 
time-dependent mass term, which deforms the ground state of canonical modes 
so that it corresponds to the Bunch-Davis vacuum during the dS phase. 
The continuity of the wave function uniquely determines the state during the 
subsequent RD phase as well. The state is 
Gaussian at all times, as is evident from eq. \eqref{solschrod}, 
even though the coefficient of the quadratic term in the exponent is \emph{complex}. 
This implies that the tracing out of degrees of freedom can be carried out 
via the evaluation of Gaussian integrals, as in the original calculation.

The state of the overall system reads
\begin{equation}
\Psi \left( \mathbf{{x}} \right)  \sim 
 \exp \left( - \frac{1}{2}  \mathbf{{x}}^T {W} \,
 \mathbf{{x}}  \right)  ,
 \label{Overall_State}
\end{equation}
where the vector $\mathbf{{x}}$ denotes collectively the 
field values $f_{lm,j}$, following the notation of \cite{srednicki,Boutivas,Katsinis}, 
as we explained in the previous subsection. 
The matrix $W$ is given by
\begin{equation}
W = \frac{\Omega}{b^2(\tau,\Omega)}-i\frac{b'(\tau,\Omega)}{b(\tau,\Omega)} .
\label{matrixW}
\end{equation}
The matrix $\Omega$ is the eigenfrequency matrix that corresponds to the time-independent part of the Hamiltonian of the overall system. It is the positive square root of the matrix of couplings $K$ between the degrees of freedom. In other words, if $x_i$ are the coordinates and $\pi_i$ the conjugate momenta, the time-independent part of the Hamiltonian reads
\begin{equation}
	H = \frac{1}{2} \sum_i \pi_i^2 + \frac{1}{2} \sum_i \sum_j x_i K_{ij} x_j .
\end{equation}
As is evident by equation (\ref{eq:Hamiltonian_discretized}), the Hamiltonian can be divided into angular momentum sectors that do not interact, allowing for the writing of the matrix $K$ as
	$K = \bigotimes\limits_{l , m} \, K_l $,
where
\begin{equation}
	\left( K_l \right)_{ij} = \left( 2 + \frac{l \left( l + 1 \right) + 1 / 2}{i^2} \right) \delta_{ij} - \frac{\left( i + \frac{1}{2} \right)^2}{i \left( i + 1 \right)} \delta_{i + 1 , j} - \frac{\left( j + \frac{1}{2} \right)^2}{j \left( j + 1 \right)} \delta_{i , j + 1} .
\end{equation}
The matrices $K_l$ have positive eigenvalues $k_{l i}$. Each $N \times N$ matrix $K_l$ has $2^N$ square roots, that have eigenvalues $\pm \sqrt{k_{l i}}$. As positive square root $\Omega_l$ of the matrix $K_l$ we define its square root that has only positive eigenvalues. Then, the matrix $\Omega$ reads
	$\Omega = \bigotimes\limits_{l , m} \, \Omega_l$.
%In short, the matrix $\Omega$ is the eigenfrequency matrix of the overall system. 
In eq. (\ref{matrixW})
we have
indicated explicitly the dependence of the function $b(\tau)$ of the previous subsection
on the eigenvalues of this matrix. 
The matrix $W$ is in general a complex symmetric matrix. 
The overall system is described by the density matrix
\begin{equation}
\rho \left( \mathbf{x} ; \mathbf{x}^\prime \right) \sim
 \exp \left[ - \frac{1}{2}  \left( 
 \mathbf{{x}}^T W\, \mathbf{{x}}  + \mathbf{{x}}^{\prime T} W^*\, \mathbf{{x}}'  
 \right)  \right],
\end{equation}
which is Gaussian.

We consider as subsystem $1$ the set of $n$ degrees of freedom 
described by the coordinates $x_j$, where $j \leq n$. 
The rest of the degrees of freedom comprise the complementary subsystem 2. 
We trace out subsystem 2.
%\footnote{Even though our notation, which follows \cite{srednicki}, indicates that we trace out the interior of the sphere, the numerical calculation is performed by actually tracing out the exterior. Moreover, we make sure that the volume of the interior is always smaller than that of the exterior, so that the former corresponds to the smaller subsystem. In any case, the result is independent of the choice of the part that is traced out, since the overall system lies in a pure state.}
It is convenient to introduce the block notation
\begin{equation}
W = \left( \begin{array}{cc} A & B \\ B^T & C \end{array} \right) , \quad\quad \mathbf{x} = \left( \begin{array}{c} \mathbf{x}_1 \\ \mathbf{x}_2 \end{array} \right)  ,
\label{block-form}
\end{equation}
where, in an obvious manner, the matrix $A$ is an $n \times n$ matrix, 
the vector $\mathbf{x}_1$ is an $n$-dimensional vector, and so on. 
The matrices $A$ and $C$ are complex symmetric matrices like $W$, 
whereas the matrix $B$ has no specific symmetry property and is not even a square matrix. 
It is a matter of simple algebra with Gaussian integrals to show that the reduced density matrix describing subsystem 1 assumes the form
\begin{equation}
\rho_1(\bx_1,\bx'_1) \sim \exp \left( -\frac{1}{2}\bx_1^T\gamma\, \bx_1  -\frac{1}{2}\bx^{\prime T}_1\gamma\, \bx'_1 +\bx^{\prime T}_1 \beta\, \bx_1
+\frac{i}{2}\bx_1^T\delta\, \bx_1  -\frac{i}{2}\bx^{\prime T}_1\delta \, \bx'_1
\right),
\label{reduced_density_matrix}
\end{equation}
where
\begin{align}
\gamma -i \delta &= A - \frac{1}{2} B {\rm Re} \left( C \right)^{-1} B^T , 
\label{gamma} \\
\beta &= \frac{1}{2} B^* {\rm Re} \left( C \right)^{-1} B^T .
\label{beta}
\end{align}

In order to calculate the entanglement entropy, one needs to specify the eigenvalues of the above reduced density matrix. In order to do so, we must solve for its eigenfunctions,
which satisfy
\begin{equation}
\int d^n\, \mathbf{x}_1^\prime\, {\rho}_1 \left( \mathbf{x}_1 ; \mathbf{x}_1^\prime
 \right) {f} \left( \mathbf{x}_1^\prime \right) = \lambda {f} \left( \mathbf{x}_1 \right) .
\end{equation}
There are two differences between our case and the simpler case of a field theory at its ground
state in flat space \cite{srednicki}. 
The first one is that the matrix $\delta$ does not vanish. 
This is a very innocent change, as it can be shown that the matrix $\delta$ does not alter the eigenvalues of the reduced density matrix, but only its eigenfunctions \cite{Boutivas,Katsinis}. 
Therefore, for the purpose of the specification of the spectrum of the reduced density matrix, the matrix $\delta$ can be set to zero by hand.

The second difference induces several complications. The matrix $\beta$ is a complex Hermitian matrix, whereas in the original calculation \cite{srednicki} it is real and symmetric. If these
properties persisted in our case, the matrices $\gamma$ and $\beta$ would be simultaneously diagonalizable via an appropriate coordinate transformation, resulting in a reduced density matrix that could be factored into the tensor product of density matrices describing a single degree of freedom each. Each of these density matrices would describe a simple harmonic oscillator mode at a thermal state, with temperatures given as functions of the eigenvalues of the matrix $\gamma^{-1} \beta$.
In our case this is not possible and the calculation is more complicated. Nevertheless, it can be shown that the general structure of the eigenfunctions remains identical. There is still a `ground' eigenstate, $n$ `first excited' eigenstates and so on. However, the reduced density matrix cannot be factored into density matrices of a single degree of freedom each.

The complete analysis of the problem is presented in refs. \cite{Boutivas,Katsinis}, 
to which the reader is referred for the details.
As it turns out, the eigenvalues of the reduced density matrix can be specified in terms of the matrix ${\cal W}$ that satisfies the quadratic matrix equation
\begin{equation}
{\cal W} = I - \tilde{\beta}^T \left( I + {\cal W} \right)^{- 1} \tilde{\beta} ,
\label{eq:exponent_eigenfunctions}
\end{equation}
where
\begin{equation}
\tilde{\beta} = \gamma^{-\frac{1}{2}} \beta \gamma^{-\frac{1}{2}} .
\label{beta_tilde}
\end{equation}
This matrix appears in the exponent of the Gaussian part of the eigenfunctions of the reduced density matrix. For example, the `ground' Gaussian eigenstate reads
\begin{equation}
\Psi_0 \left( \mathbf{x} \right) \sim \exp \left( - \frac{1}{2} \mathbf{x}^T {\cal W} \mathbf{x} \right) .
\label{eq:eigenstate_ground}
\end{equation}
Then, the eigenvalues of the reduced density matrix assume the form
\begin{equation}
\lambda_{\left\{ m_1 , m_2 , \ldots , m_n \right\}} = \left( 1 - \xi_1 \right) \left( 1 - \xi_2 \right) \ldots \left( 1 - \xi_{N-n} \right) \xi_1^{m_1} \xi_2^{m_2} \ldots \xi_n^{m_n} ,
\label{eq:spectrum_reduced_eigenvalues}
\end{equation}
where $m_i$, $i = 1 , 2 , \ldots , n$, take any non-negative integer value and the parameters $\xi_i$ are the eigenvalues of the matrix
\begin{equation}
\Xi = \tilde{\beta}^T \left( I + {\cal W} \right)^{- 1} .
\label{eq:spectrum_Xi_def}
\end{equation}
Notice that the matrix $\Xi$ is neither real or Hermitian, yet it turns out that 
it possesses real eigenvalues. The reason is that it is similar to a Hermitian matrix,
as shown in section 4.4.2 of ref. \cite{Katsinis}. 

A complication that appears in this calculation is that the matrix equation \eqref{eq:exponent_eigenfunctions} has more than one solutions. Only one of them gives rise to normalizable eigenfunctions of the reduced density matrix. This is the only one which,
through eq. \eqref{eq:spectrum_Xi_def}, corresponds to a matrix $\Xi$ that has no eigenvalues larger than 1, and thus to an appropriately normalized spectrum of the reduced density matrix \eqref{eq:spectrum_reduced_eigenvalues}. The problem of selecting the correct solution 
of eq. \eqref{eq:exponent_eigenfunctions}
can be bypassed by upgrading it to the eigenvalue problem of a matrix of higher dimension. 
More specifically, it can be shown that the eigenvalues of the matrix $\Xi$ are a subset of the eigenvalues of the matrix
\begin{equation}
M=
\begin{pmatrix}
2 \hat{\beta}^{-1} & -\hat{\beta}^{-1}\hat{\beta}^T\\
I & 0
\end{pmatrix}.
\label{eq:M_def}
\end{equation}
These eigenvalues $\lambda$ are the roots of the characteristic polynomial
\begin{equation}
\det \left(2 I - \lambda \hat{\beta} - \frac{1}{\lambda} \hat{\beta}^T \right) = 0 .
\label{eq:spectrum_eigenvalues_Xi_equation}
\end{equation}
Because of its structure, the eigenvalues come in pairs, with one element being the inverse of the other. Thus, half of the eigenvalues are smaller than 1 and the rest are larger than 1. 
The eigenvalues of the matrix $\Xi$ that corresponds to the correct ${\cal W}$ are exactly the eigenvalues of $M$ that are smaller than 1. This allows the specification of the spectrum of the reduced density matrix in a way that is amenable to a numerical calculation. 
The entanglement entropy is directly given by the formula
\begin{equation}
S_{\mathrm{EE}} = - \sum_i \left( \ln \left( 1 - \xi_i \right) + \frac{\xi_i}{1 - \xi_i} \ln \xi_i \right) .
\label{eq:spectrum_SEE}
\end{equation}
The reader is referred to \cite{Boutivas,Katsinis} for the proof of the above statements
and more details.

\section{Numerical results}\label{numerics}

\subsection{General expectations}

The formalism presented in the previous section provides the basis for the numerical 
calculation of the entanglement entropy associated with a massless free field in an expanding
background. 
As we discussed in the introduction, we assume the presence of an UV cutoff 
$k_s$ for the wavenumbers of the modes that contribute to the entropy. For 
a discretized theory on a lattice, this cutoff is set by the lattice 
spacing $\epsilon$, so that $k_s\sim 1/\epsilon$. 
An IR cutoff $k_l$ may also be present, even though we find that it does not
have an effect on the entropy.
For the discretized theory, the value of $k_l$ is set by the total size of the lattice.  
The structure of the theory implies that
the result for the entropy depends on the combination $H \epsilon$, with $H$ the constant 
value of the Hubble parameter during the dS era, which also sets the initial value of this
parameter in the subsequent RD era. Our assumption that the relevant physical 
degrees of freedom
are the ones that exit the horizon during inflation and 
re-enter during the RD era implies that $H \epsilon \sim 1$ 
(for a scale factor set equal to 1 at the dS-RD transition). 

Before presenting the numerical results, it is instructive to consider the
expected form of the entanglement entropy and its dependence on the various
scales of the theory. In the dS phase the leading contributions to the entropy, apart from 
a constant, are expected to be 
	\begin{equation}
		S=c_1\frac{A_p}{\epsilon_p^2}+\frac{ c_2}{2} \log \left(\frac{A_p}{\epsilon_p^2} \right)
		+c_4  \log(H \epsilon_p) H^2 A_p 
		+c_5H^2 A_p + \frac{c_6}{2}\log(H^2 A_p ).
		\label{expansion}
		\end{equation}
We have used physical parameters and followed the notation of \cite{pimentel}, neglecting
a term involving the mass of the field.
Here $A_p=4\pi  a^2(\tau) R^2$ is the proper area of the entangling surface and
$\epsilon_p=a(\tau) \epsilon$ the physical UV cutoff. The terms proportional to $c_1$ 
and $c_2$ are present in a flat background as well. The coefficient $c_1$ is 
regularization-scheme dependent and was first computed in the seminal work of 
Srednicki \cite{srednicki}, while the universal constant $c_2=-1/90$ was computed in 
\cite{casini0,Lohmayer}. The term proportional to $c_4$ is an additional UV divergence
appearing in a curved background. The terms proportional to $c_5$ and $c_6$ are 
genuine IR effects related to the expansion. The value $c_6=1/90$ was found in
\cite{pimentel}.

The numerical accuracy of our calculation allows the precise determination of only
the leading area terms. The logarithmic terms proportional to $c_2$ and $c_6$ cannot be reliably specified within the accuracy of our numerical calculations. The logarithmic dependence of the
term proportional to $c_4$ affects the coefficient of the area law and is tractable. 
We postpone its precise analysis for a
future publication. It must be pointed out that no volume term is expected to 
develop in the dS phase, despite the fact that the modes lie in squeezed states. 
The technical reasons in the context of our approach 
are explained in section 5 of \cite{Boutivas}.

The form of the entanglement entropy in the RD phase is more complicated. Several
curvature invariants may contribute, which are reduced to the terms involving $H$ in
the dS phase. Similarly to before, an area term is present. However, as we shall
see, a new feature appears: a volume term develops and becomes the dominant 
contribution to the entropy. A particular question, which we address in 
subsection \ref{volumesection}, is whether this term arises from UV contributions,
or it is a low-energy effect.

\subsection{The evolution of the entropy}

\begin{figure}[t!]
	\centering
	\includegraphics[width=1.0\textwidth]{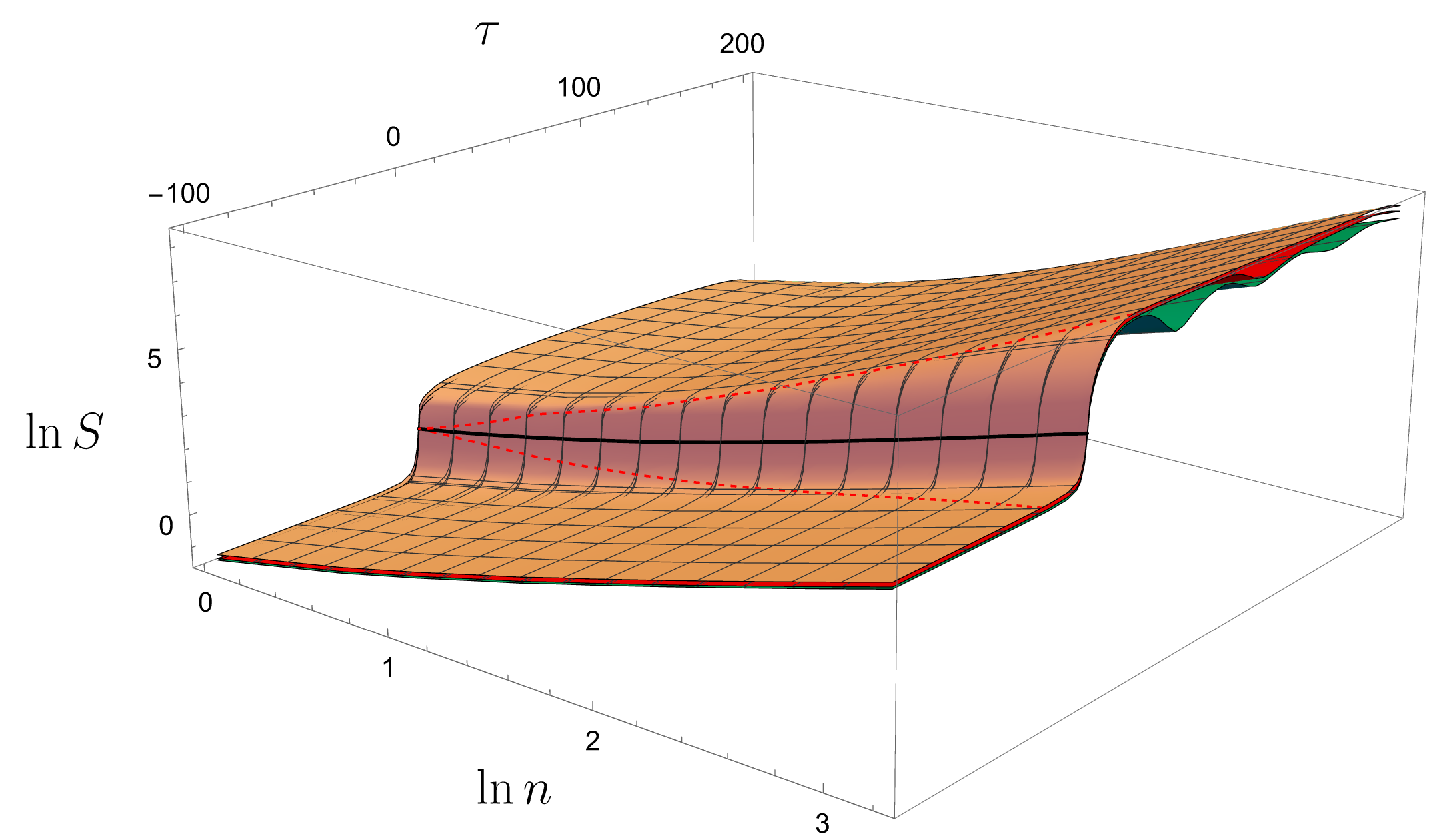} 
	\caption{ 
		The entanglement entropy for a spherical 
		region as a function of the radius of the entangling surface and time for $H\epsilon =1$. 
		The radius of the total
		spherical lattice is $L=N \epsilon$, and we depict the results for
		$N=200$ (brown surface), $N=100$ (red surface), and $N=50$ (green surface). 
		We also indicate the entropy at the dS to RD transition (black curve) and the
		location of the comoving horizon (dashed, red curve).
	}
	\label{fig1}
\end{figure}

In fig. \ref{fig1} we depict the entanglement entropy for a spherical 
region as a function of the radius of the entangling surface and time for $H\epsilon =1$. 
The entropy and radius axes are logarithmic. 
The discretization of the theory and the implementation
of the UV cutoff were described in subsection \ref{discrHam}. The radius of the total
spherical lattice is $L=N \epsilon$, and we depict the results for
$N=200$ (brown surface), $N=100$ (red surface), $N=50$ (green surface). 
We have displaced slightly the three surfaces in the vertical direction in order to
make them more visible. 
The radius of the entangling surface is $R=n\epsilon$, with $n=1,...,25$, so that it 
is always smaller that half of the radius of the overall lattice and the interior region contains manifestly fewer degrees of freedom than the exterior.
The transition from the dS to the RD era takes place at $\tau=0$, in a region 
characterized by a strong increase of the entanglement entropy. 
The thick black line depicts the entropy as a function of $n$ at $\tau=0$.
We have also indicated the location of the comoving horizon at various times through a dashed, red
line. For $\tau<0$ in the dS era, the comoving horizon shrinks and a diminishing
part of the total system 
remains subhorizon. For $\tau>0$ in the RD era, the comoving horizon grows and the subhorizon part of the
system grows. At the dS to RD transition only a small part, with size of the order of the
lattice spacing, is subhorizon.

We observe that, for $\tau \to -\infty$, the entropy becomes independent of time. 
In the logarithmic plot, $\ln S$ is a linear function of $\ln n$ with slope equal to 2
to a very good approximation, apart from expected deviations for small $n$, resulting
from subleading contributions to the entropy (constant and logarithmic terms). We have 
ignored such corrections, as the precision of our calculation is not sufficient to 
determine them reliably, and we have focused on large values of $n$ for which terms 
of quadratic or of higher order dominate. For $\tau \to -\infty$, 
we have performed a fit of the entropy
with a quadratic function in the region $n\geq 15$ in order to determine the coefficient $s$
of the quadratic term.  
We obtain $s\simeq 0.09$ for all three values of $N$. 
This is to be compared with the value $s \simeq 0.3$ quoted by Srednicki \cite{srednicki}.
The difference arises from the different regularization scheme that we use, as we
discussed in detail in subsection \ref{discrHam}. In order to confirm the reliability  
of our numerics, we have also computed the entropy using the regularization of \cite{srednicki},
reproducing the value $s\simeq 0.3$.
The form of the entropy is consistent with the expectation that it should coincide with 
the entropy in a static background when the {\it physical} radius of the entangling surface is
much smaller than the Hubble radius. This is the case for $\tau \to -\infty$, 
when the scale factor approaches zero and the term proportional to 
$c_1$ in eq. (\ref{expansion}) dominates. This result is also consistent with the 
assumption of a Bunch-Davies vacuum.

When the time approaches $\tau=0$, there is a strong increase of the
entanglement entropy while the background is still in the dS phase. The dependence on
$n$ remains strictly quadratic until the transition to the RD era. This feature
is expected for a free theory, as was discussed in \cite{Boutivas}.
The increase of the entropy arises because the terms proportional to 
$c_4$ and $c_5$ in eq. (\ref{expansion}) become dominant. In particular, we have
verified that the entropy grows $\sim a^2(\tau)$, following the growth of the 
physical radius of the entangling surface when the comoving radius is kept fixed,
as in our calculation. 
The transition to the RD phase is followed by a further increase of the entropy,
associated with the additional squeezing of the wave function of the canonical modes.
For $\tau \to +\infty$ the entanglement entropy develops an almost constant form, with
a profile that indicates deviations from a purely quadratic dependence on $n$. We
investigate this form in more detail below. What is apparent in fig. \ref{fig1} is a
weak oscillatory behaviour that depends on $N$. In particular, the wavelength of the
oscillatory pattern is comparable to that of the longest mode allowed by the finite
lattice. This pattern is a finite-size effect that would be absent 
in an infinite system. It is also apparent that the oscillations decay with time, 
with the resulting asymptotic form of the entropy being independent of $N$. 
We have analyzed numerically the behaviour of the entropy at late times, but we have not
identified a significant residual effect associated with an IR cutoff $k_l\sim 1/L$. 
If such an effect exists, it is 
not visible with the numerical precision of our calculation.

\begin{figure}[t!]
	\centering
	\includegraphics[width=1.0\textwidth]{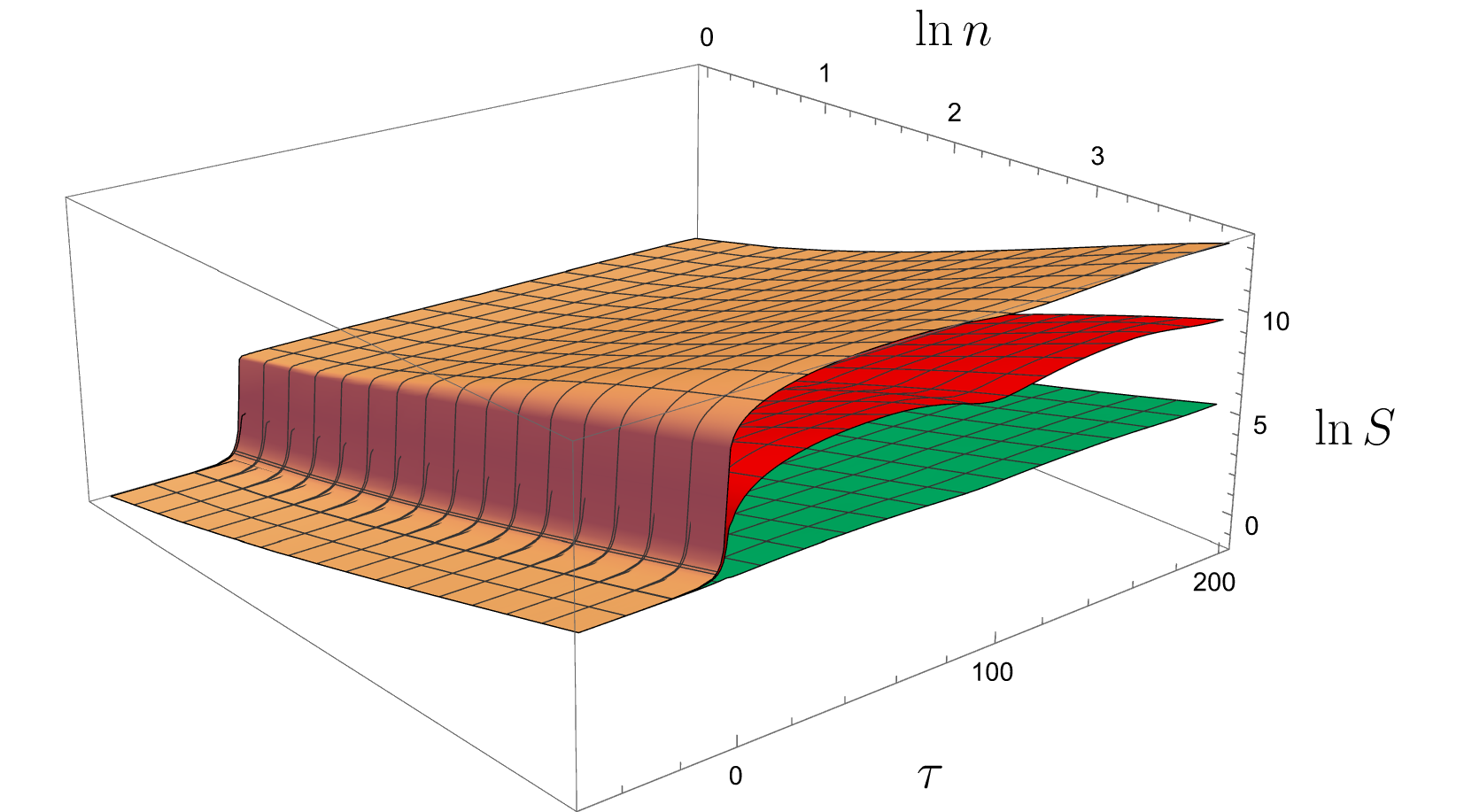} 
	\caption{ 
		The entropy as a function of the radius of the entangling surface and time for 
		$H\epsilon =0.1$ (green surface), $H\epsilon =1$ (red surface) and 
		$H\epsilon =10$ (brown surface). The size of the lattice is $N=100$.  
	}
	\label{fig2}
\end{figure}

The next issue that we would like to analyze is the dependence of the entanglement
entropy on the Hubble scale $H$ during inflation. As we have already mentioned, the
relevant dimensionless parameter is the product $H\epsilon$. 
The value of $s$ we quoted corresponds to the coefficient of the term 
$n^2=R^2/\epsilon^2=(a R)^2/(a \epsilon)^2$ that involves the ratio of the physical
radius to the physical UV cutoff. 
In fig. \ref{fig2} we depict
the form of the entropy as a function of the radius and time for 
$H\epsilon =0.1$ (green surface), $H\epsilon =1$ (red surface) and 
$H\epsilon =10$ (brown surface). The size of the lattice is $N=100$.  

During the dS era, if the UV cutoff $\epsilon$
is much shorter than the Hubble radius $1/H$, as in the case $H\epsilon=0.1$, 
the entropy is dominated by the short-distance
entanglement between degrees of freedom on either side of the entangling surface. The 
expansion of the background induces only a subleading effect even when the physical entangling
surface is larger than the horizon. During the subsequent RD era, the physical cutoff 
shrinks even further relatively to the Hubble radius, so that the entropy continues to
be dominated by the UV, similarly to the situation in static space.

For larger values of $H\epsilon$ during inflation, the asymptotic entanglement entropy
of the discretized theory at late times becomes larger than that for a static background. 
As we discussed in the introduction, we focus on the case $H \epsilon=1$, for which
modes that did not cross the horizon during inflation and never froze 
are not taken into account in the calculation of the entropy. For this choice of UV cutoff,
the enhancement of the entropy due to the expansion of the background
can be significant relatively to the static-space case, 
as can be seen in fig. \ref{fig2}, in which the
vertical axis is logarithmic. Notice that the physical UV cutoff today, which is 
of the order of $1\rm{m}$ as we discussed in the introduction, is much shorter than the Hubble
radius today. However, the effect of the squeezing of the wave function during inflation and
the transition to the RD era results in an increase of the entropy by at least two orders of
magnitude relatively to what one would expect for a static background. 
Since the determination of the shortest wavelength that froze at the end of
inflation is rather imprecise, we have also considered the case $H \epsilon=10$,
for which the enhancement of the entropy relatively to the static case is 
higher by an additional two orders of magnitude.

\subsection{The volume term} \label{volumesection}

\begin{figure}[t!]
\centering
\includegraphics[width=0.48\textwidth]{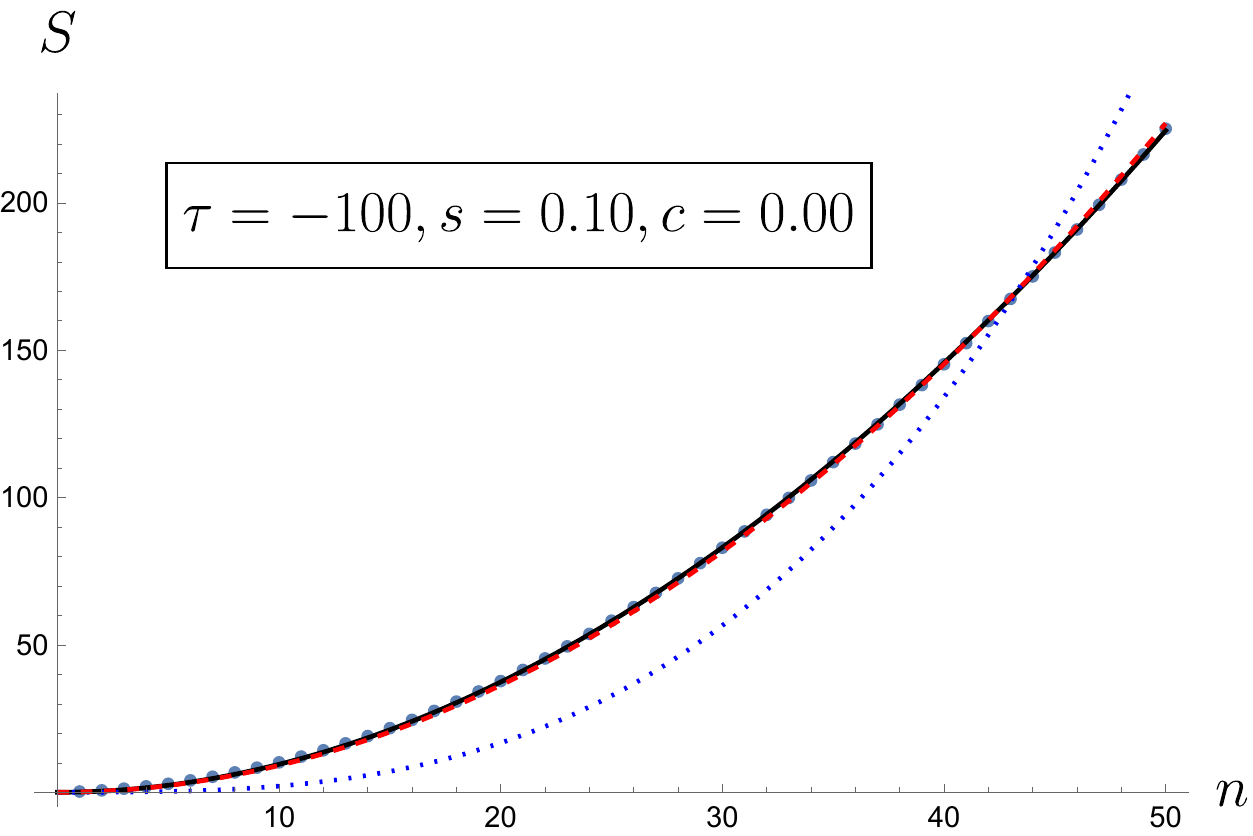} 
\includegraphics[width=0.48\textwidth]{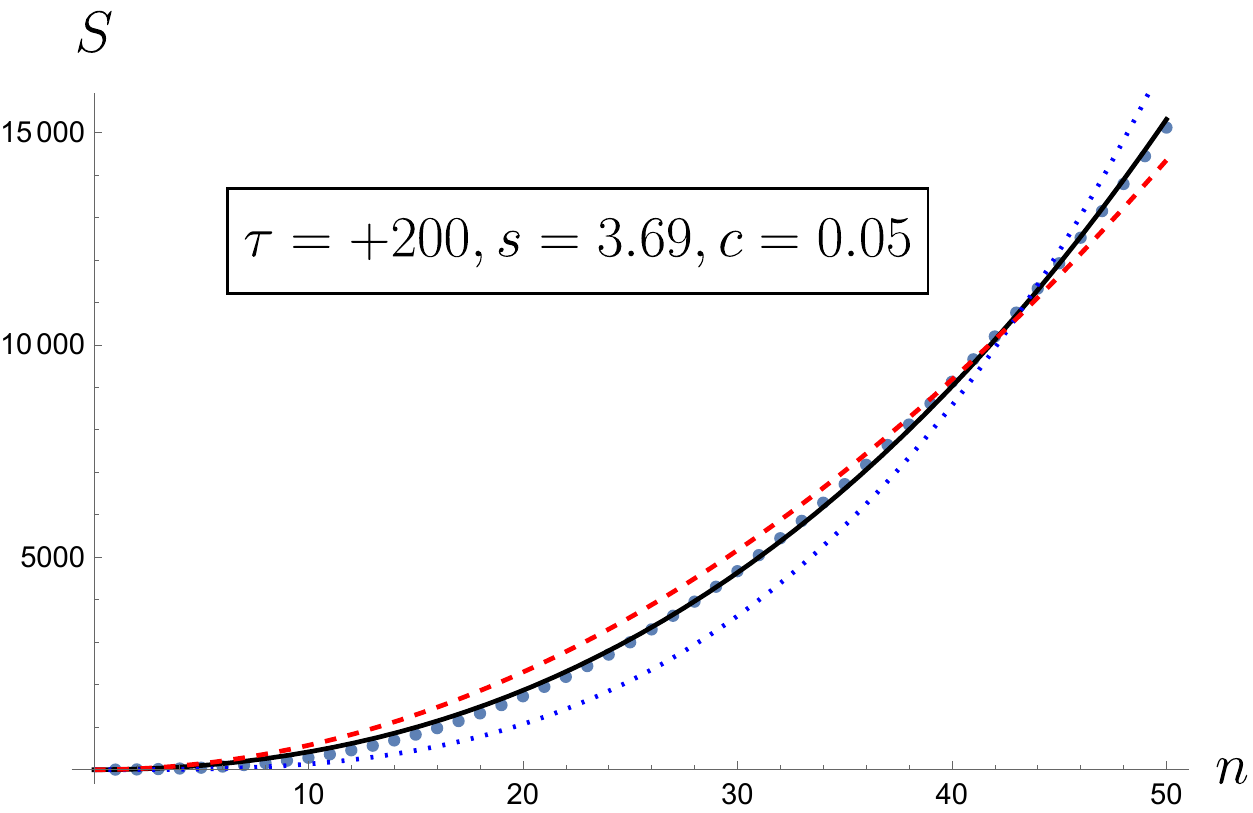} 
\caption{
The entanglement entropy as a function of $n=R/\epsilon$
at an early time $\tau/\epsilon=-100$ in the dS era and a late time $\tau/\epsilon=+200$
in the RD era. 
The size of the lattice is $L=N \epsilon$ with $N=100$.
The red, dashed
curve and the blue, dotted one correspond to fits with only a quadratic or cubic term.
}
\label{fig3}
\end{figure}

Apart from the enhancement of the entanglement entropy, a very interesting feature that 
is apparent in figs. \ref{fig1} and \ref{fig2} is the dependence of the entropy 
on the radius of the entangling surface
at late times. As long as the background is in the dS phase, the entropy is dominated by a quadratic term. However, after the transition to the RD era, a different pattern emerges. 
In fig. \ref{fig3} we depict the entanglement entropy as a function of $n=R/\epsilon$
at an early time $\tau/\epsilon=-100$ in the dS era and a late time $\tau/\epsilon=+200$
in the RD era.
The size of the lattice is $L=N \epsilon$ with $N=100$. 
We fit the shape of the curve for $n\geq 15$ with
a polynomial that includes a quadratic and a cubic term. The assumed form of the entropy
is 
\begin{equation}
	S=s \frac{R^2}{\epsilon^2}+c \frac{R^3}{\epsilon^3},
\label{fite} \end{equation}	
where we have indicated explicitly the use of the UV cutoff in the parameterization. The coefficients $s$ and $c$ are in general functions of $\tau / \epsilon$.
Their values at specific times are given within the inset. At early times $c$ vanishes,
while at late times a good fit is obtained only if it takes a nonvanishing value. 
The red, dashed curve and the blue, dotted one correspond to fits with only a quadratic
or cubic term, respectively. It is clear that at late times they both are unsatisfactory.

\begin{figure}[t!]
	\centering
	\includegraphics[width=0.48\textwidth]{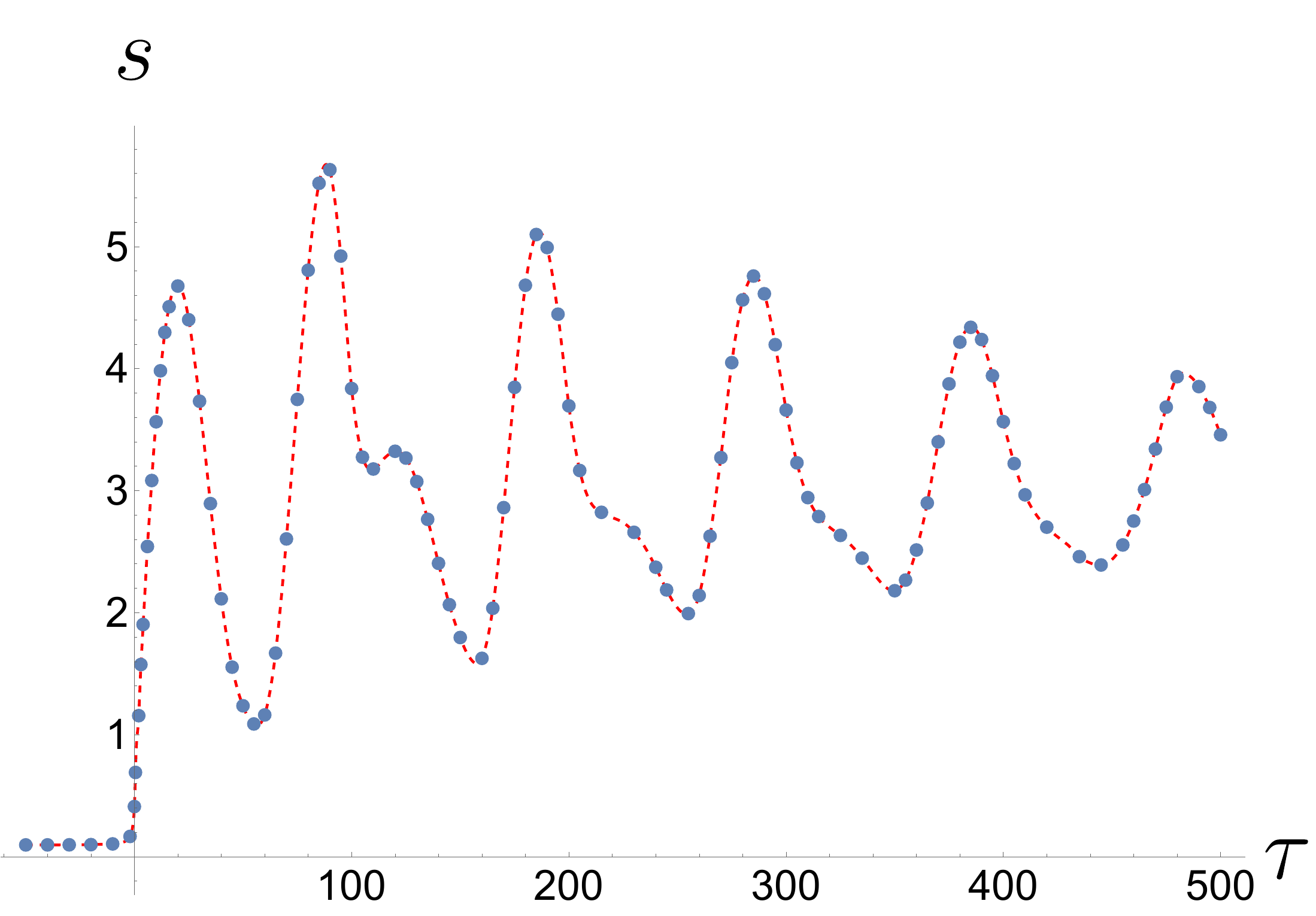} 
	\includegraphics[width=0.48\textwidth]{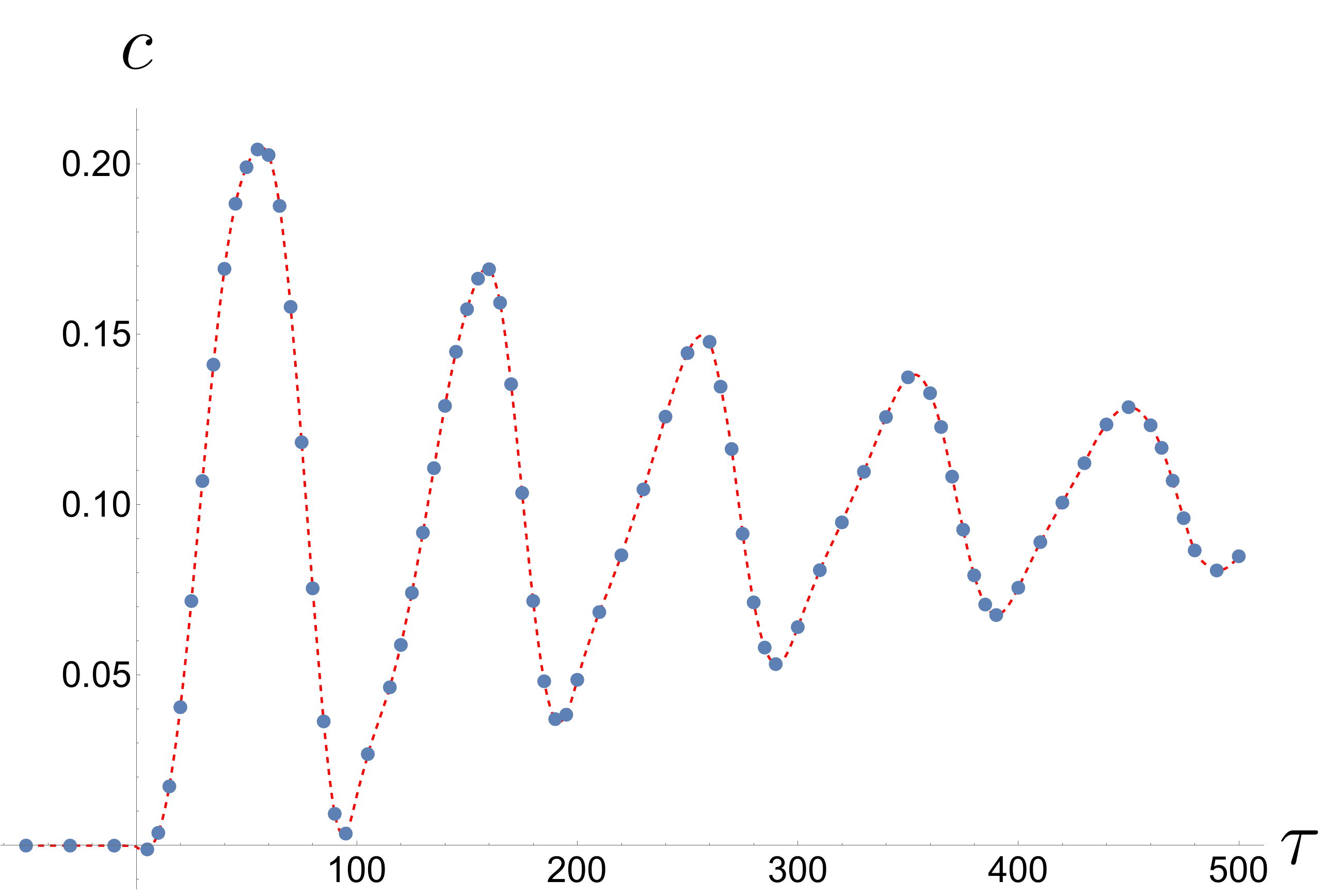} 
	\caption{
		The coefficients $s$ (left plot) and $c$ (right plot)
		of the quadratic and the cubic term, respectively, as a function of time, for the theory
		of fig. \ref{fig3}.
	}
	\label{fig4}
\end{figure}

In fig. \ref{fig4} we depict the coefficents $s$ (left plot) and $c$ (right plot)
of the quadratic and the cubic term, respectively, as a function of time.
 At early times
we have $s\simeq 0.09$ and $c=0$. While the background is still in the dS phase, $s$
grows, whereas $c$ remains zero. The cubic term emerges only after the time $\tau=0$
of the dS to RD transition. The oscillatory patterns arise because of the 
finite-size effect that we discussed earlier in this section, 
which is associated with the longest
mode that can exist within the finite lattice. The oscillations decay with time and 
the two parameters asymptotically settle to constant values: $s\simeq 2.5$ and $c\simeq 0.13$.   
The growth of the cubic term is a very interesting, novel feature. It indicates that 
strong entanglement is not confined only to narrow regions on either side of the 
entangling surface, but spreads throughout the whole bulk of the system. 
We discuss the implications of this result in the following section.

\begin{figure}[t!]
	\centering
	\includegraphics[width=0.48\textwidth]{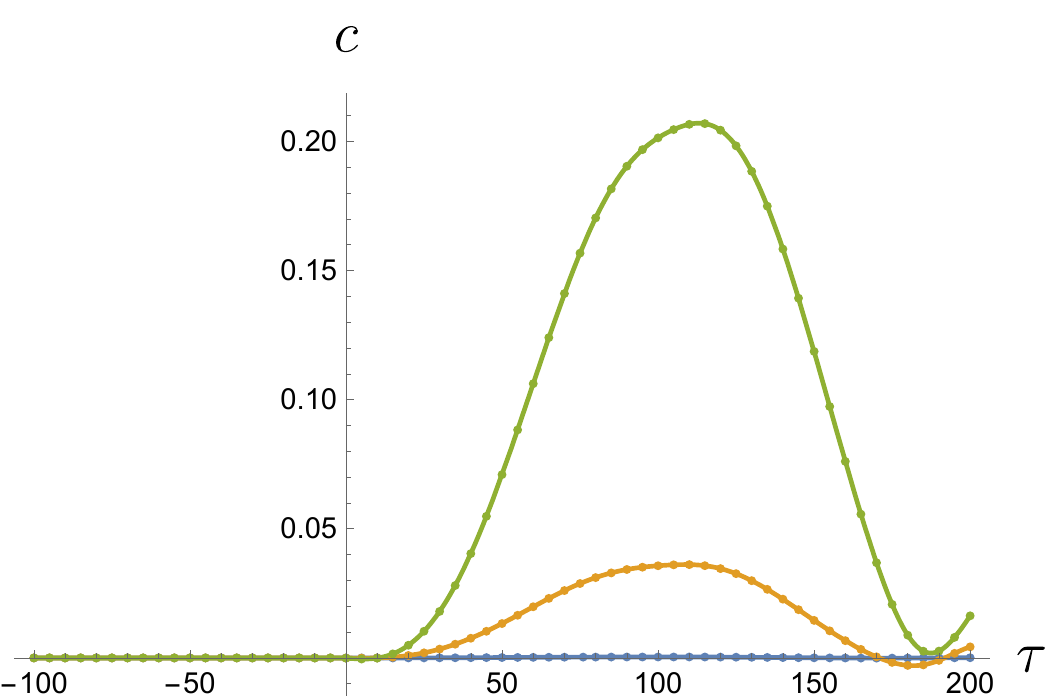} 
	\includegraphics[width=0.48\textwidth]{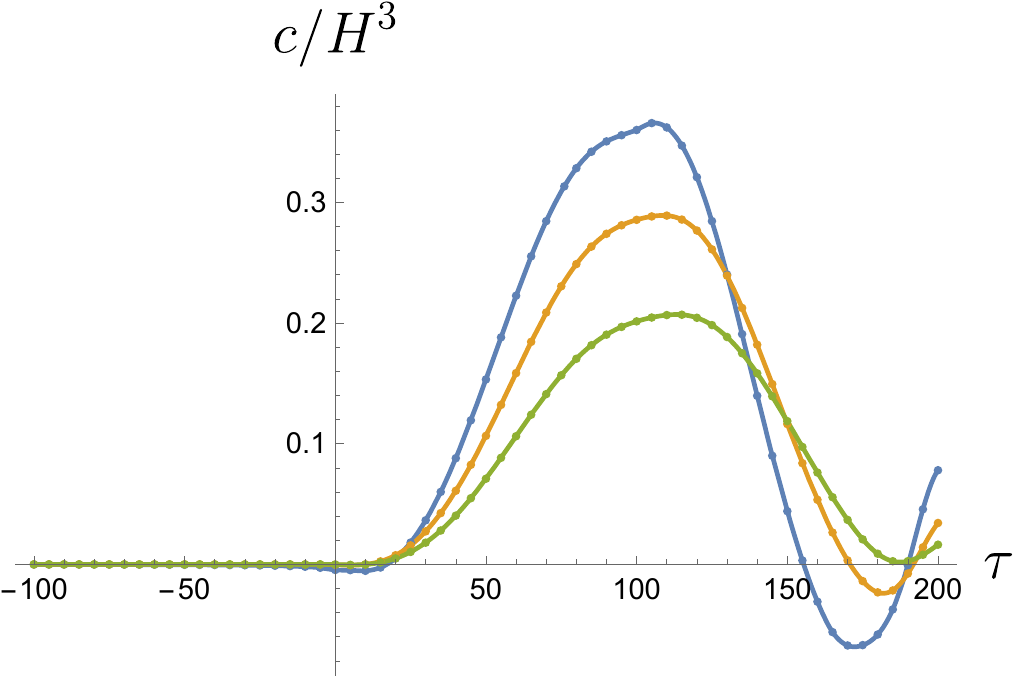} 
	\caption{
		The coefficients $c$ (left plot) and $c/(H\epsilon)^3$ (right plot) of 
		the cubic term, as a function of time, for 
		$H\epsilon=0.1$ (blue curve), $H\epsilon=0.5$ (orange curve), and $H\epsilon=1$ (green curve). The size of the lattice is $L=N \epsilon$ with $N=200$.
		We have set $\epsilon=1$.
	}
	\label{fig5}
\end{figure}

An important question is whether the appearance of the volume term is an effect 
related to the UV modes of the system. The parameterization (\ref{fite}) gives the
impression that the dependence on the UV cutoff is strong. However, 
adopting the logic that led to eq. (\ref{expansion}) suggests a parameterization 
of the form 
\begin{equation}
S=\tilde{s} H^2 R_p^2+\tilde{c} H^3R_p^3,
	\label{fith} \end{equation}	
with $R_p=a(\tau)R$ the physical radius. 
Apart from the effects related to the expansion,
the area term receives contributions from the UV range.
However, this is not 
necessarily true for the cubic term. Comparison of eqs. (\ref{fite}), (\ref{fith})
gives
\begin{equation}
		\tilde{c}=\frac{c}{(H\epsilon)^3a^3(\tau)}.
	\label{scsc}
\end{equation}
The explicit UV dependence has now been shifted into the definition of $\tilde{c}$.
However, this could be counterbalanced by the dependence of $c$ on $H\epsilon$.
In fig. \ref{fig5} we depict $c$ and $c/(H\epsilon)^3$ for 
$H\epsilon=0.1$ (blue curve), $H\epsilon=0.5$ (orange curve), and $H\epsilon=1$ (green curve). Notice that the size of the lattice is $L=N \epsilon$ with $N=200$, i.e.
twice that in fig. \ref{fig4}. This explains the larger period of the 
oscillations associated with the longest mode of the system (a finite-size effect discussed in the previous subsection). 
The slightly negative values of $c$ within a short time interval are caused by the
limitations of the numerically calculation and the fit to the results. 

The important conclusion that can be drawn from fig. \ref{fig5} is that the 
UV dependence of $c$ is such that $\tilde{c}$ is largely independent of $\epsilon$.
Even though $c$ varies by 3 orders of magnitude for the values of $H\epsilon$ we 
considered, the average value of $c/(H\epsilon)^3$ varies by less than a factor of 2. 
The other important point is that the volume term is determined by the comoving
radius $R=R_p/a(\tau)$. When this term dominates, the entanglement entropy within
a spherical region that follows the expansion is roughly constant. 

The qualitative picture that emerges is that the entanglement entropy within 
a spherical region of fixed comoving radius $R$ grows during the dS era, even for a 
constant comoving UV cutoff $\epsilon$. This growth is caused by the squeezing of the
wave function of the various field modes by the expansion,
and is additive to the standard entanglement entropy in a static background.
However, the dependence on the radius is quadratic, so that the effect satisfies an
area law. The transition to the RD phase induces the emergence of a volume term
in the entropy. The most plausible interpretation is that the entanglement is 
spread by the expansion over larger distances, until it encompasses the degrees of 
freedom of the whole system. The volume contribution to the entropy is roughly
constant for a fixed comoving radius. In this sense, the expansion during the
RD phase does not generate additional entropy, but redistributes over the
whole system the entropy produced during the dS phase.

\section{Conclusions}\label{conclusions}

We have followed the evolution of the entanglement entropy of a real, massless, scalar field 
throughout the inflationary period 
and the subsequent era of radiation domination. We have assumed that during inflation 
the field is in the Bunch-Davies vacuum. As a result, the entanglement 
of the short-distance modes is very similar to that at the ground state 
in a static background. For longer modes,
the transition towards a 
squeezed state upon horizon exit of each mode during inflation and 
the additional squeezing when radiation domination sets in enhance the entanglement entropy.
Even though we did not analyse it explicitly, a similar behaviour is expected during 
matter domination, as has been verified in a previous study \cite{Boutivas}.

The entanglement entropy, even in a static background, is an UV-divergent quantity
that requires the introduction of an UV cutoff. 
We have identified this cutoff with the wavelength $\lambda_s$ 
of the last mode that exited the horizon at the 
end of inflation and re-entered immediately in the RD era. Modes with shorter wavelengths 
did not go through the process of freezing 
and the dominance of the growing term in the mode function. We have adopted the view that
they can be considered as vacuum fluctuations even today, while we focused on the entanglement 
due to modes that are
directly accessible to observations. 

A question of particular interest concerns the dependence of the entropy on the size of
the system. The characteristic pattern in a static
$(3+1)$-dimensional background is that the entropy is
proportional to the area of the entangling surface, similarly to the black hole entropy
\cite{srednicki}.
This feature is attributed to the strong entanglement of 
short-distance modes on the two sides of the entangling surface. One would expect
that the rapid expansion during the
dS era would stretch the entangled modes sufficiently far from the entangling surface, so that
a volume effect might appear. However, this expectation is not realized. The subtle technical
reasons are explained in section 5 of \cite{Boutivas}. On the other hand, the expansion during
the RD era induces a drastic modification. A volume term develops and becomes the 
dominant contribution to the entropy at late times. It must be noted that the
presence of a volume term is not unexpected for systems lying in squeezed states, such as
those developing through the expansion of the background \cite{Katsinis}. It is rather a peculiarity of the
cosmological evolution that the volume term appears only after the universe has
evolved into the RD era.

The behaviour that we outlined above must be contrasted with the standard 
picture of quantum to classical transition upon horizon exit 
for the cosmological fluctuations produced during inflation. 
In our analysis we have fully accounted for the quantum properties of the field,
without discarding the decaying mode, no matter how smaller it may become than the growing one.
In this sense, the entropy that we have computed should be attributed to the quantum
entanglement of the degrees of freedom. For the short-distance modes it coincides
with the entanglement entropy in a static background.

At late times during the RD era, the volume contribution  
to the entanglement entropy within a fixed comoving radius
approaches a constant value. 
A similar behaviour is expected during matter domination.
The possibility that the quantum properties of a massless field can
be traced even today seems very exciting. 
Even though we studied a free scalar field, 
our analysis could be relevant for all very weakly interacting fields as, for example, 
gravitational waves resulting from tensor modes during inflation, 
under the assumption that they evolve almost freely and their quantum coherence 
is not lost during the whole evolution of the universe until today through some secondary
process.  
Finding appropriate experimental signatures of the quantum origin of such fields
seems a difficult task \cite{maldacenabell,venninbell,vennin3}. 
However, the question is of fundamental importance \cite{micheligw,domcke2}.

Even though the picture outlined above seems the most natural, 
the final form of the entropy, and in particular the 
appearance of a volume term, points towards a possibly different interpretation.
It is known that the effective decoherence underlying the  
quantum to classical transition is achieved
only if the decaying mode is neglected. On the other hand, 
it has also been pointed out that even 
an exponentially small decaying mode may be important for the calculation of the
entropy of the perturbations \cite{classical1}.
Moreover, there is no clear criterion that can characterize the
system as classical when the decaying mode is not neglected and the state remains 
pure \cite{morikawa,quantumness,chandran}. 
It is then possible that the smallness of the decaying term results in a system
with significant entropy that must be interpreted in classical terms. 
Our result emerges through the non-trivial form of
the density matrix induced by a rapidly expanding background that contains a horizon. 
A possible interpretation would attribute thermal characteristics to the
reduced density matrix resulting from tracing out degrees of freedom beyond a certain radius, 
which are classically inaccessible to an observer. The regions beyond the horizon provide the typical example for such a situation. The presence of the volume term would then 
support the interpretation of the entropy 
of the observable universe as thermal, 
even though its origin would lie in the presence of the horizon.

Let us recall the simple example of two coupled harmonic oscillators lying at their ground state \cite{srednicki}. The reduced density matrix that describes either of the two is the density matrix of a single harmonic oscillator\footnote{The eigenfrequency of this effective oscillator is equal to the geometric mean of the eigenfrequencies of the two canonical modes.}, lying at a mixed state, which is thermal with a temperature that depends on the coupling between the two oscillators. 
An observer with knowledge of the existence of both oscillators concludes that the form of this reduced density matrix results from  
the entanglement between them and the corresponding entropy is entanglement entropy. However, an observer who has access only to one of the 
two oscillators cannot reach such a conclusion. According to this observer, the only way to interpret the entropy is as thermal.

In our study we calculated the entanglement entropy in the expanding universe splitting the system into two through a spherical entangling surface. A realization of this surface
is provided by the cosmological horizon, which physically prohibits an observer from measuring the observables associated to the degrees of freedom in the exterior. A hypothetical observer with
knowledge of the entire universe would attribute
the entropy to the entanglement between the interior and the exterior. However, an observer confined within the horizon can only perceive the entropy as thermal. 
The presence of the volume term that we deduced implies that this constrained observer does not just see entropy emanating from the cosmological horizon, similarly to a black hole,
but concludes that the bulk of the interior has been heated. This conclusion is reached despite
the fact that the whole system still lies at a pure state.

One must bear in mind that in our case the reduced density matrix does not 
correspond to a system in exact thermal equilibrium. 
When a subsystem of a harmonic system contains many degrees of freedom, 
the reduced density matrix is organized through effective canonical modes, each of 
which lies at a different temperature. 
When squeezing is present this picture is not exact, since these canonical modes cannot be associated with a real combination of the local degrees of freedom. However, the appearance of many temperatures (as many as the degrees of freedom of the subsystem) is still valid. 
This fact is a consequence of the integrability of our toy model, which is a free field theory. 
The thermalization hypothesis suggests that in a realistic theory the reduced density matrix would actually be thermal.

We note that there is no distinct point in the evolution that we observed 
which can be associated
with a qualitative change in the nature of the entropy. The evolution is smooth, as can
be seen in figs. \ref{fig1} and \ref{fig2}. The structure of the reduced density matrix
is complicated and does not provide any hints for a qualitative transition.
On the other hand, the thermodynamic interpretation
is appealing because it is consistent with the quantum to classical transition. The quantum effects are 
tied to the decaying mode, which can be many orders of magnitude smaller than the
growing one for realistic scenarios of inflation. 
A thermal entropy is consistent with the picture of a stochastic classical field,
as a result of the dominance of the growing mode.

The thermodynamic interpretation could provide a 
quantum-mechanical realization of reheating after inflation. In our analysis the
background is introduced by hand and not as a solution of the Einstein equations
for an appropriate equation of state. In this sense, the RD era does not arise as
a result of the field properties. On the other hand, it is intriguing that the 
appearance of a volume term, a characteristic feature of thermal entropy, 
is connected to the transition to the RD era. 

It is usually assumed that the reheating occurs through the strong interactions between
the products of the inflaton decay. Interpreting our result as thermal entropy would 
imply  
that all fields (interacting or not) that exist at the end of inflation
can be treated in the same way,
since they would all become part of the thermal environment during reheating. 
Of course, there is no indication of a common temperature for all the constituents of the
universe. However, our understanding of the situation is still very incomplete.

Our numerical results indicate that the magnitude of the entropy can be significant. 
If we estimate it through the volume term that develops during the era of 
radiation domination, we get a value for the observable universe 
$\sim (H \lambda_s)^{-3}$, with a cutoff $\lambda_s \sim 1\rm{m}$, as we discussed
in the introduction, while the current value of the Hubble radius
is $1/H \sim 10^{26}\rm{m}$. This gives a value
$\sim 10^{78}$ for the entropy, which
is to be compared with the standard thermal entropy  
$\sim 10^{88}$ associated with
the plasma in the early universe, transferred to the photons and neutrinos
today. 
There is a discrepancy of 10 orders of magnitude, which can be 
understood if we recall that the typical wavelength of the cosmic background photons is
$\sim 10^{-3}\rm{m}$. The origin of the difference lies in that the scale of the 
standard thermal
entropy is connected to the energy density of the inflaton background, while the 
scale of the entropy
we computed for a massless field is set by the last mode that exited the horizon at the
end of inflation. The two would be comparable if the energy scale of inflation was close to
the Planck scale. However, the observations indicate an energy scale roughly three orders
of magnitude lower. 
Despite the quantitative discrepancy, the interpretation of the entropy as thermal
does not seem completely off the mark. 

The basic conclusion that can be reached through our analysis is that the rapid 
expansion of the background during inflation and the transition to the RD era  
generate a significant amount of entropy even for a system of 
a non-interacting field. This entropy is carried by the long-distance modes 
that go through the process of horizon exit during inflation and are accessible to 
experiments today. It can be defined and computed 
as entanglement entropy, and is intrinsically linked to the presence of a horizon. 
It seems unlikely that the possible loss of quantum 
coherence, which we did not address in this work, will result in the suppression
of the entropy. It seems more likely that the reduced density matrix, 
employed by an observer with no access to the regions beyond the horizon, will
remain nontrivial, with a thermodynamic interpretation. 
This speculation is consistent with the standard picture of the quantum to 
classical transition during inflation.    
More work is needed in order to obtain a deeper understanding of these issues. 
The notion that the entropy of the universe
can be attributed to the presence of the cosmological horizon merits further exploration.

\acknowledgments

N.Tetradis would like to thank
I. Dalianis, V. Domcke, S. Mathur, K. Papadodimas, M. Simonovic for very useful discussions. 
The research of K. Boutivas and N. Tetradis was supported by the Hellenic Foundation for
Research and Innovation (H.F.R.I.) under the “First Call for H.F.R.I.
Research Projects to support Faculty members and Researchers and
the procurement of high-cost research equipment grant” (Project
Number: 824). The research of D.Katsinis was supported by the 
FAPESP Grant No. 2021/01819-0.


\begin{thebibliography}{99}

\bibitem{inflation1}
V.~F.~Mukhanov and G.~V.~Chibisov,
``Quantum Fluctuations and a Nonsingular Universe,''
JETP Lett. \textbf{33} (1981), 532-535.

\bibitem{inflation2}
S.~W.~Hawking,
``The Development of Irregularities in a Single Bubble Inflationary Universe,''
Phys. Lett. B \textbf{115} (1982), 295.
%doi:10.1016/0370-2693(82)90373-2

\bibitem{inflation3}
A.~H.~Guth and S.~Y.~Pi,
``Fluctuations in the New Inflationary Universe,''
Phys. Rev. Lett. \textbf{49} (1982), 1110-1113.
%doi:10.1103/PhysRevLett.49.1110

\bibitem{inflation4}
A.~A.~Starobinsky,
``Dynamics of Phase Transition in the New Inflationary Universe Scenario and Generation of Perturbations,''
Phys. Lett. B \textbf{117} (1982), 175-178.
%doi:10.1016/0370-2693(82)90541-X

\bibitem{inflation5}
J.~M.~Bardeen, P.~J.~Steinhardt and M.~S.~Turner,
``Spontaneous Creation of Almost Scale - Free Density Perturbations in an Inflationary Universe,''
Phys. Rev. D \textbf{28} (1983), 679.
%doi:10.1103/PhysRevD.28.679

\bibitem{physrep}
V.~F.~Mukhanov, H.~Feldman and R.~H.~Brandenberger,
``Theory of cosmological perturbations. Part 1. Classical perturbations. Part 2. Quantum theory of perturbations. Part 3. Extensions,''
Phys. Rept. \textbf{215} (1992), 203-333.
%doi:10.1016/0370-1573(92)90044-Z

\bibitem{maldacenabell}
J.~Maldacena,
``A model with cosmological Bell inequalities,''
Fortsch. Phys. \textbf{64} (2016), 10-23
doi:10.1002/prop.201500097
[arXiv:1508.01082 [hep-th]].

\bibitem{venninbell}
J.~Martin and V.~Vennin,
``Obstructions to Bell CMB Experiments,''
Phys. Rev. D \textbf{96} (2017) no.6, 063501
doi:10.1103/PhysRevD.96.063501
[arXiv:1706.05001 [astro-ph.CO]].


   
   \bibitem{vennin3}
   L.~Espinosa-Portal\'es and V.~Vennin,
   ``Real-space Bell inequalities in de~Sitter,''
   JCAP \textbf{07} (2022) no.07, 037
   %doi:10.1088/1475-7516/2022/07/037
   [arXiv:2203.03505 [quant-ph]].
   
   
   
   
   \bibitem{momentumspace1}
   V.~Balasubramanian, M.~B.~McDermott and M.~Van Raamsdonk,
   ``Momentum-space entanglement and renormalization in quantum field theory,''
   Phys. Rev. D \textbf{86} (2012), 045014
   %doi:10.1103/PhysRevD.86.045014
   [arXiv:1108.3568 [hep-th]].
   
   \bibitem{momentumspace2}
   S.~Brahma, O.~Alaryani and R.~Brandenberger,
   ``Entanglement entropy of cosmological perturbations,''
   Phys. Rev. D \textbf{102} (2020) no.4, 043529
   %doi:10.1103/PhysRevD.102.043529
   [arXiv:2005.09688 [hep-th]].
   
   \bibitem{momentumspace3}
   S.~Brahma, A.~Berera and J.~Calder\'on-Figueroa,
   ``Quantum corrections to the primordial tensor spectrum: open EFTs \& Markovian decoupling of UV modes,''
   JHEP \textbf{08} (2022), 225
   %doi:10.1007/JHEP08(2022)225
   [arXiv:2206.05797 [hep-th]].
   
   \bibitem{momentumspace4}
   S.~Brahma, J.~Calder\'on-Figueroa, M.~Hassan and X.~Mi,
   ``Momentum-space entanglement entropy in de Sitter spacetime,''
   Phys. Rev. D \textbf{108} (2023) no.4, 043522
   %doi:10.1103/PhysRevD.108.043522
   [arXiv:2302.13894 [hep-th]].
  
  
  \bibitem{tripathy}
  S.~Tripathy, R.~N.~Raveendran, K.~Parattu and L.~Sriramkumar,
  ``Amplifying quantum discord during inflationary magnetogenesis through violation of parity,''
  Phys. Rev. D \textbf{108} (2023) no.12, 123512
%  doi:10.1103/PhysRevD.108.123512
  [arXiv:2306.16168 [gr-qc]].
 
 %\cite{Bombelli:1986rw}
 \bibitem{Bombelli:1986rw}
 L.~Bombelli, R.~K.~Koul, J.~Lee and R.~D.~Sorkin,
 ``A Quantum Source of Entropy for Black Holes,''
 Phys. Rev. D \textbf{34}, 373-383 (1986).
 %doi:10.1103/PhysRevD.34.373
 %1206 citations counted in INSPIRE as of 07 Nov 2022
 
 \bibitem{srednicki}
 M.~Srednicki,
 ``Entropy and area,''
 Phys. Rev. Lett. \textbf{71} (1993), 666-669
 %doi:10.1103/PhysRevLett.71.666
 [arXiv:hep-th/9303048 [hep-th]]. 
  
  \bibitem{callan}
  C.~G.~Callan, Jr. and F.~Wilczek,
  ``On geometric entropy,''
  Phys. Lett. B \textbf{333} (1994), 55-61
 % doi:10.1016/0370-2693(94)91007-3
  [arXiv:hep-th/9401072 [hep-th]].
  
  
     \bibitem{wilczek}
     C.~Holzhey, F.~Larsen and F.~Wilczek,
     ``Geometric and renormalized entropy in conformal field theory,''
     Nucl. Phys. B \textbf{424} (1994), 443-467
  %   doi:10.1016/0550-3213(94)90402-2
     [arXiv:hep-th/9403108 [hep-th]].
     
     
  \bibitem{muller}
  R.~Muller and C.~O.~Lousto,
  ``Entanglement entropy in curved space-times with event horizons,''
  Phys. Rev. D \textbf{52} (1995), 4512-4517
  %doi:10.1103/PhysRevD.52.4512
  [arXiv:gr-qc/9504049 [gr-qc]].
  
  
  \bibitem{korepin}
  V.~E.~Korepin,
  ``Universality of Entropy Scaling in One Dimensional Gapless Models,''
  Phys. Rev. Lett. \textbf{92} (2004), 096402
  %doi:10.1103/PhysRevLett.92.096402
  [arXiv:cond-mat/0311056 [cond-mat.str-el]].   
     
  
  \bibitem{cardy1}
  P.~Calabrese and J.~L.~Cardy,
  ``Entanglement entropy and quantum field theory,''
  J. Stat. Mech. \textbf{0406} (2004), P06002
  %doi:10.1088/1742-5468/2004/06/P06002
  [arXiv:hep-th/0405152 [hep-th]].   
     
  \bibitem{cardy2}   
  P.~Calabrese and J.~Cardy,
  ``Entanglement entropy and conformal field theory,''
  J. Phys. A \textbf{42} (2009), 504005
  %doi:10.1088/1751-8113/42/50/504005
  [arXiv:0905.4013 [cond-mat.stat-mech]].   
      
  
  \bibitem{casini0}
  H.~Casini and M.~Huerta,
  ``Entanglement entropy in free quantum field theory'',
  J. Phys. A \textbf{42}, 504007 (2009)
  %doi:10.1088/1751-8113/42/50/504007
  [arXiv:0905.2562 [hep-th]].
  %555 citations counted in INSPIRE as of 29 Nov 2022
  
    
    \bibitem{Lohmayer}
    R.~Lohmayer, H.~Neuberger, A.~Schwimmer and S.~Theisen,
    ``Numerical determination of entanglement entropy for a sphere,''
    Phys. Lett. B \textbf{685} (2010), 222-227
    %doi:10.1016/j.physletb.2010.01.053
    [arXiv:0911.4283 [hep-lat]].
  
  \bibitem{casini1}
  H.~Casini and M.~Huerta,
  ``Entanglement entropy for the n-sphere,''
  Phys. Lett. B \textbf{694} (2011), 167-171
  %doi:10.1016/j.physletb.2010.09.054
  [arXiv:1007.1813 [hep-th]].
  
  \bibitem{casini2}
    H.~Casini, M.~Huerta and R.~C.~Myers,
    ``Towards a derivation of holographic entanglement entropy,''
    JHEP {\bf 1105} (2011) 036
  %  doi:10.1007/JHEP05(2011)036
    [arXiv:1102.0440 [hep-th]].
  
  \bibitem{pimentel}
  J.~Maldacena and G.~L.~Pimentel,
  ``Entanglement entropy in de Sitter space,''
  JHEP \textbf{02} (2013), 038
  %doi:10.1007/JHEP02(2013)038
  [arXiv:1210.7244 [hep-th]].
  
  \bibitem{Kanno:2014lma}
  S.~Kanno, J.~Murugan, J.~P.~Shock and J.~Soda,
  ``Entanglement entropy of $\alpha$-vacua in de Sitter space,''
  JHEP \textbf{07} (2014), 072
  %doi:10.1007/JHEP07(2014)072
  [arXiv:1404.6815 [hep-th]].
  
  \bibitem{Iizuka:2014rua}
  N.~Iizuka, T.~Noumi and N.~Ogawa,
  ``Entanglement entropy of de Sitter space $\alpha$-vacua,''
  Nucl. Phys. B \textbf{910} (2016), 23-29
  %doi:10.1016/j.nuclphysb.2016.06.024
  [arXiv:1404.7487 [hep-th]].
  
  \bibitem{Kanno:2016qcc}
  S.~Kanno, M.~Sasaki and T.~Tanaka,
  ``Vacuum State of the Dirac Field in de Sitter Space and Entanglement Entropy,''
  JHEP \textbf{03} (2017), 068
  %doi:10.1007/JHEP03(2017)068
  [arXiv:1612.08954 [hep-th]].
  
  \bibitem{stefan}
  J.~Berges, S.~Floerchinger and R.~Venugopalan,
  ``Dynamics of entanglement in expanding quantum fields,''
  JHEP \textbf{04} (2018), 145
  %doi:10.1007/JHEP04(2018)145
  [arXiv:1712.09362 [hep-th]].
  

  
     \bibitem{colas}
     T.~Colas, J.~Grain and V.~Vennin,
     ``Four-mode squeezed states: two-field quantum systems and the symplectic group $\mathrm {Sp}(4,{\mathbb {R}})$,''
     Eur. Phys. J. C \textbf{82} (2022) no.1, 6
     %doi:10.1140/epjc/s10052-021-09922-y
     [arXiv:2104.14942 [quant-ph]].
     
     \bibitem{vennin1}
     J.~Martin and V.~Vennin,
     ``Real-space entanglement of quantum fields,''
     Phys. Rev. D \textbf{104} (2021) no.8, 085012
     %doi:10.1103/PhysRevD.104.085012
     [arXiv:2106.14575 [hep-th]].
  
   
     \bibitem{vennin2}
     J.~Martin and V.~Vennin,
     ``Real-space entanglement in the Cosmic Microwave Background,''
     JCAP \textbf{10} (2021), 036
     %doi:10.1088/1475-7516/2021/10/036
     [arXiv:2106.15100 [gr-qc]].
  
  \bibitem{Boutivas}
  K.~Boutivas, G.~Pastras and N.~Tetradis,
  ``Entanglement and expansion,''
  JHEP \textbf{05} (2023), 199
%  doi:10.1007/JHEP05(2023)199
  [arXiv:2302.14666 [hep-th]].
  



\bibitem{Katsinis}
D.~Katsinis, G.~Pastras and N.~Tetradis,
``Entanglement of harmonic systems in squeezed states,''
JHEP \textbf{10} (2023), 039
%doi:10.1007/JHEP10(2023)039
[arXiv:2304.04241 [hep-th]].





\bibitem{ryu1}
  S.~Ryu and T.~Takayanagi,
  ``Holographic derivation of entanglement entropy from AdS/CFT,''
  Phys.\ Rev.\ Lett.\  {\bf 96} (2006) 181602
%  doi:10.1103/PhysRevLett.96.181602
  [hep-th/0603001].
  
  \bibitem{ryu2}
    T.~Nishioka, S.~Ryu and T.~Takayanagi,
    ``Holographic Entanglement Entropy: An Overview,''
    J.\ Phys.\ A {\bf 42} (2009) 504008
 %   doi:10.1088/1751-8113/42/50/504008
    [arXiv:0905.0932 [hep-th]].
    
\bibitem{review1}
  S.~Ryu and T.~Takayanagi,
  ``Aspects of Holographic Entanglement Entropy,''
  JHEP {\bf 0608} (2006) 045
%  doi:10.1088/1126-6708/2006/08/045
  [hep-th/0605073].
  


\bibitem{adscft1}
  J.~M.~Maldacena,
  ``The Large N limit of superconformal field theories and supergravity,''
  Int.\ J.\ Theor.\ Phys.\  {\bf 38} (1999) 1113
   [Adv.\ Theor.\ Math.\ Phys.\  {\bf 2} (1998) 231]
%  doi:10.1023/A:1026654312961, 10.4310/ATMP.1998.v2.n2.a1
  [hep-th/9711200].
 
 \bibitem{adscft2}
  S.~S.~Gubser, I.~R.~Klebanov and A.~M.~Polyakov,
  ``Gauge theory correlators from noncritical string theory,''
  Phys.\ Lett.\ B {\bf 428} (1998) 105
%  doi:10.1016/S0370-2693(98)00377-3
  [hep-th/9802109].
 
 \bibitem{adscft3}
  E.~Witten,
  ``Anti-de Sitter space and holography,''
  Adv.\ Theor.\ Math.\ Phys.\  {\bf 2} (1998) 253
%  doi:10.4310/ATMP.1998.v2.n2.a2
  [hep-th/9802150].

\bibitem{tetradisgiataganas}
D.~Giataganas and N.~Tetradis,
``Entanglement entropy in FRW backgrounds,''
Phys. Lett. B \textbf{820} (2021), 136493
%doi:10.1016/j.physletb.2021.136493
[arXiv:2105.12614 [hep-th]].


\bibitem{tetradisgiantsos}
V.~Giantsos and N.~Tetradis,
``Entanglement entropy in a four-dimensional cosmological background,''
Phys. Lett. B \textbf{833} (2022), 137331
%doi:10.1016/j.physletb.2022.137331
[arXiv:2203.06699 [hep-th]].

\bibitem{luongo1}
A.~Belfiglio, O.~Luongo and S.~Mancini,
``Inflationary entanglement,''
Phys. Rev. D \textbf{107} (2023) no.10, 103512
%doi:10.1103/PhysRevD.107.103512
[arXiv:2212.06448 [gr-qc]].

\bibitem{luongo2}
A.~Belfiglio, O.~Luongo and S.~Mancini,
``Entanglement area law violation from field-curvature coupling,''
Phys. Lett. B \textbf{848} (2024), 138398
%doi:10.1016/j.physletb.2023.138398
[arXiv:2306.08357 [gr-qc]].

\bibitem{Grishchuk}
L.~P.~Grishchuk and Y.~V.~Sidorov,
``Squeezed quantum states of relic gravitons and primordial density fluctuations,''
Phys. Rev. D \textbf{42} (1990), 3413-3421.
%doi:10.1103/PhysRevD.42.3413

\bibitem{squeeze1}
R.~H.~Brandenberger, T.~Prokopec and V.~F.~Mukhanov,
``The Entropy of the gravitational field,''
Phys. Rev. D \textbf{48} (1993), 2443-2455
%doi:10.1103/PhysRevD.48.2443
[arXiv:gr-qc/9208009 [gr-qc]].

\bibitem{squeeze2}
R.~H.~Brandenberger, V.~F.~Mukhanov and T.~Prokopec,
``Entropy of a classical stochastic field and cosmological perturbations,''
Phys. Rev. Lett. \textbf{69} (1992), 3606-3609
%doi:10.1103/PhysRevLett.69.3606
[arXiv:astro-ph/9206005 [astro-ph]].

\bibitem{squeeze3}
T.~Prokopec,
``Entropy of the squeezed vacuum,''
Class. Quant. Grav. \textbf{10} (1993), 2295-2306.
%doi:10.1088/0264-9381/10/11/012

\bibitem{squeeze4}
A.~L.~Matacz,
``The Coherent state representation of quantum fluctuations in the early universe,''
Phys. Rev. D \textbf{49} (1994), 788-798
%doi:10.1103/PhysRevD.49.788
[arXiv:gr-qc/9212008 [gr-qc]].

\bibitem{squeeze5}
M.~Gasperini and M.~Giovannini,
``Entropy production in the cosmological amplification of the vacuum fluctuations,''
Phys. Lett. B \textbf{301} (1993), 334-338
%doi:10.1016/0370-2693(93)91159-K
[arXiv:gr-qc/9301010 [gr-qc]].

\bibitem{squeeze6}
M.~Gasperini and M.~Giovannini,
``Quantum squeezing and cosmological entropy production,''
Class. Quant. Grav. \textbf{10} (1993), L133-L136
%doi:10.1088/0264-9381/10/9/004
[arXiv:gr-qc/9307024 [gr-qc]].

\bibitem{squeeze7}
C.~Kiefer, D.~Polarski and A.~A.~Starobinsky,
``Entropy of gravitons produced in the early universe,''
Phys. Rev. D \textbf{62} (2000), 043518
%doi:10.1103/PhysRevD.62.043518
[arXiv:gr-qc/9910065 [gr-qc]].

\bibitem{squeeze8}
D.~Campo and R.~Parentani,
``Decoherence and entropy of primordial fluctuations. I: Formalism and interpretation,''
Phys. Rev. D \textbf{78} (2008), 065044
%doi:10.1103/PhysRevD.78.065044
[arXiv:0805.0548 [hep-th]].


\bibitem{albrecht}
A.~Albrecht, P.~Ferreira, M.~Joyce and T.~Prokopec,
``Inflation and squeezed quantum states,''
Phys. Rev. D \textbf{50} (1994), 4807-4820
%doi:10.1103/PhysRevD.50.4807
[arXiv:astro-ph/9303001 [astro-ph]].

\bibitem{classical1}
D.~Polarski and A.~A.~Starobinsky,
``Semiclassicality and decoherence of cosmological perturbations,''
Class. Quant. Grav. \textbf{13} (1996), 377-392
%doi:10.1088/0264-9381/13/3/006
[arXiv:gr-qc/9504030 [gr-qc]].

\bibitem{classical2}
C.~Kiefer, D.~Polarski and A.~A.~Starobinsky,
``Quantum to classical transition for fluctuations in the early universe,''
Int. J. Mod. Phys. D \textbf{7} (1998), 455-462
%doi:10.1142/S0218271898000292
[arXiv:gr-qc/9802003 [gr-qc]].

\bibitem{classical3}
C.~Kiefer, J.~Lesgourgues, D.~Polarski and A.~A.~Starobinsky,
``The Coherence of primordial fluctuations produced during inflation,''
Class. Quant. Grav. \textbf{15} (1998), L67-L72
%doi:10.1088/0264-9381/15/10/002
[arXiv:gr-qc/9806066 [gr-qc]].

\bibitem{classical4}
B.~Allen, E.~E.~Flanagan and M.~A.~Papa,
``Is the squeezing of relic gravitational waves produced by inflation detectable?,''
Phys. Rev. D \textbf{61} (2000), 024024
%doi:10.1103/PhysRevD.61.024024
[arXiv:gr-qc/9906054 [gr-qc]].

\bibitem{classical5}
C.~Kiefer and D.~Polarski,
``Why do cosmological perturbations look classical to us?,''
Adv. Sci. Lett. \textbf{2} (2009), 164-173
%doi:10.1166/asl.2009.1023
[arXiv:0810.0087 [astro-ph]].

\bibitem{domcke}
N.~Aggarwal, O.~D.~Aguiar, A.~Bauswein, G.~Cella, S.~Clesse, A.~M.~Cruise, V.~Domcke, D.~G.~Figueroa, A.~Geraci and M.~Goryachev, \textit{et al.}
``Challenges and opportunities of gravitational-wave searches at MHz to GHz frequencies,''
Living Rev. Rel. \textbf{24} (2021) no.1, 4
%doi:10.1007/s41114-021-00032-5
[arXiv:2011.12414 [gr-qc]].



\bibitem{guerrero}
J.~Guerrero and F.~F.~L\'opez-Ruiz,
``On the Lewis\textendash{}Riesenfeld (Dodonov\textendash{}Man\textquoteright{}ko) invariant method,''
Phys. Scripta \textbf{90} (2015) no.7, 074046
%doi:10.1088/0031-8949/90/7/074046
[arXiv:1503.01371 [quant-ph]].

\bibitem{lewis1}
H.~R.~Lewis,
``Class of exact invariants for classical and quantum time-dependent harmonic oscillators,''
J. Math. Phys. \textbf{9} (1968), 1976-1986.
%doi:10.1063/1.1664532

\bibitem{lewis2}
H.~R.~Lewis and W.~B.~Riesenfeld,
``An Exact quantum theory of the time dependent harmonic oscillator and of a charged particle time dependent electromagnetic field,''
J. Math. Phys. \textbf{10} (1969), 1458-1473.
%doi:10.1063/1.1664991


\bibitem{micheligw}
A.~Micheli and P.~Peter,
``Quantum cosmological gravitational waves?,''
[arXiv:2211.00182 [gr-qc]].

\bibitem{domcke2}
D.~Carney, V.~Domcke and N.~L.~Rodd,
``Graviton detection and the quantization of gravity,''
[arXiv:2308.12988 [hep-th]].


\bibitem{morikawa}
M.~Morikawa,
``Quantum Decoherence and Classical Correlation in Quantum Mechanics,''
Phys. Rev. D \textbf{42} (1990), 2929-2932
%doi:10.1103/PhysRevD.42.2929

\bibitem{quantumness}
J.~Martin, A.~Micheli and V.~Vennin,
``Comparing quantumness criteria,''
EPL \textbf{142} (2023) no.1, 18001
%doi:10.1209/0295-5075/acc3be
[arXiv:2211.10114 [quant-ph]].

\bibitem{chandran}
S.~M.~Chandran, K.~Rajeev and S.~Shankaranarayanan,
``Real-space quantum-to-classical transition of time dependent background fluctuations,''
Phys. Rev. D \textbf{109} (2024) no.2, 023503
%doi:10.1103/PhysRevD.109.023503
[arXiv:2307.13611 [gr-qc]].



\end{thebibliography}
\end{document}